\documentclass[preprint,authoryear,review,12pt]{elsarticle}

\usepackage{amssymb}
\usepackage{amsmath}
\usepackage{endnotes}
\usepackage{color}
\definecolor{White}{rgb}{1,1,1}
\usepackage{lscape}

\usepackage[left=3cm, right=3cm]{geometry}

\raggedright
\def\beq{\begin{equation}}
\def\eeq#1{\label{#1}\end{equation}}
\def\beqa{\begin{eqnarray}}
\def\eeqa#1{\label{#1}\end{eqnarray}}

\usepackage[running]{lineno}

\journal{}

\begin{document}

\thispagestyle{empty}

\noindent Title:  A niche remedy for the dynamical problems of neutral theory
%A blend of niche and neutral theory  
%\vskip-0.2cm
%\indent ~quantifies the impact of competition on extinction
\\
Authors:  Andrew E. Noble$^{1,2}$ and William F. Fagan$^{1}$
\\
Affiliation:  $^1$Department of Biology, 
\vskip-0.2cm
\indent \indent ~ University of Maryland, College Park, MD 20742, USA
\\
\indent \indent ~$^2$Department of Environmental Science and Policy, 
\vskip-0.2cm
\indent \indent ~ University of California, Davis, California 95616, USA 
\\
\\
Abtract:  We demonstrate how niche theory and Hubbell's original formulation of neutral theory can be blended together into a general framework modeling the combined effects of selection, drift, speciation, and dispersal on community dynamics.  This framework connects many seemingly unrelated ecological population models, and allows for quantitative predictions to be made about the impact of niche stabilizing and destabilizing forces on population extinction times and abundance distributions.  In particular, the existence of niche stabilizing forces in our blended framework can simultaneously resolve two major problems with the dynamics of neutral theory, namely predictions of species lifetimes that are too short and species ages that are too long.  
%Abtract:  Efforts to reconcile Hubbell's neutral theory with niche theory have been limited by the lack of a nested analytical framework in which departures from the neutral hypothesis might be quantified in fits to empirical data.  Here, we present an extension of Hubbell's theory to allow for asymmetries in competitive, density-dependent interactions such that niche dynamics stabilize or destabilize coexistence over intermediate time scales, while speciation and extinction regulate an unstable coexistence over evolutionary time scales.  Intraspecific exceeding interspecific competition delays extinction beyond neutral expectations, while the opposing scenario generates a stochastic Allee effect that  accelerates extinction for rare species.  We discuss how these mechanisms, acting separately or together, might resolve the problems of short species lifetimes and long species ages in the dynamics of Hubbell's neutral theory.
\vskip-0.2cm
\indent\indent drift, speciation, dispersal, species lifetime problem, species age problem

\clearpage
\setcounter{page}{1}
%\linenumbers

%% main text
\section*{Introduction}
\label{intro} 

%Niche vs. neutral.
Niche theory and Hubbell's neutral theory~\citep{PHubbell:2001p4284} offer two distinct mechanisms for the long-term maintenance of biodiversity in competitive communities.  In niche theory, if species limit themselves more than each other, then asymmetric, density-dependent interactions stabilize coexistence, and species composition remains invariant over long time scales~\citep{Chesson:2000p5531}.  In neutral theory, functionally equivalent species drift steadily to extinction, but speciation balances extinction to maintain an unstable coexistence~\citep{Caswell:1976p6359,PHubbell:2001p4284,Chave:2004p1260}.  In simulations where the only two free parameters are speciation and migration levels, \citet{PHubbell:2001p4284} found remarkably good fits to data from various closed-canopy tree communities, the mixed mesophytic forest on the Cumberland Plateau of Kentucky, a planktonic copepod community of the northeastern Pacific gyre, and the bat community of Barro Colorado Island, among others.  However, in fits to data from forest plots along the Manu River of Amazonian Peru, abundances for the top seven species exceeded the neutral prediction.  \citet{PHubbell:2001p4284} referred to these discrepancies as ``ecological dominance deviations" and, in an extension of his simulation, found that small asymmetries in survival across species were sufficient to obtain a good fit.  This empirical evidence of asymmetries was an early indicator of the need to blend niche and neutral theory.  

%Neutral statics good.
The publication of Hubbell's book triggered a heated debate over the utility of a neutral theory for community ecology.  The debate focused largely on neutral theory predictions for RSA data~\citep{Volkov:2003p1299, McGill:2003p6201, Volkov:2005p6326, Etienne:2005p6998, Etienne:2005p6363, Chave:2006p6372, Marani:2006gu, Adler:2007p8512, Mcgill:2007p4667, Muneepeerakul:2008p2327, Mutshinda:2008p5577, Jabot:2008va, Levine:2009p6069, Volkov:2009p7487, Adler:2010p8532, Stokes:2010p9971, Ofiteru:2010p12626, Jeraldo:2012}.  An important outcome of this debate was the recognition that both niche and neutral processes can generate similar RSA distributions~\citep{Chave:2002p6357, Purves:2005, Alonso:2006p4856, Kelly:2008p10029, Chisholm:2010p9932, Noble:2011p10042}, and that niche and neutral theory might be ``two ends of a continuum" \citep[p.~179]{Chase:2003}.  Early efforts to explore that continuum in simulations \citep{Tilman:2004p4673, Gravel:2006p4668} have been followed by analytical efforts to reconcile niche and neutral theory~\citep{Marani:2006gu, Kadmon:2007p7648, Etienne:2007p6358, Walker:2007p4670, Haegeman:2008p6364, Loreau:2008vf, Allouche:2009p6355, Mutshinda:2011wf}.  The work of \citet{Haegeman:2011p9946} is particuarly relevant here.  By adding demographic stochasticity and immigration to a Lotka-Volterra model, they generated an analytical theory of local communities that combines niche and neutral dynamics.  However, because the zero-sum rule is absent from the model of~\citet{Haegeman:2011p9946}, Hubbell's original theory cannot be recovered as a limiting case.  

%Neutral dynamics not so good.
The fundamental problem with neutral theories appears to lie not in its static estimates of RSA distributions but rather in its dynamical estimates of species lifetimes~\citep{Ricklefs:2003p6692,Nee:2005p9980,Ricklefs:2006p6745,Rosindell:2010p8838,Chisholm:2014vy,Odwyer:2014je}.  The lifetime of a species in the metacommunity is commonly defined as the time period from speciation to extinction~(see, e.g.,~\citep{Chisholm:2014vy}).  Neutral theory with a point speciation mechanism~\citep{PHubbell:2001p4284} predicts expected species lifetimes -- the mean value of the species lifetime distribution~\citep{Pigolotti:2005ts} -- that are too short when compared with data.  This is the ``species lifetime" problem~\citep{Chisholm:2014vy}.  Neutral theory also predicts expected lifetimes for abundant species -- due to fat, power-law tails in the species lifetime distribution~\citep{Pigolotti:2005ts} -- that are too long.  This is the ``species age" problem~\citep{Chisholm:2014vy}.  One solution to the species lifetime problem is protracted speciation, allowing for a transition period between the origination of a cryptic species and its taxonomic recognition, as suggested by~\citep{Ricklefs:2003p6692} and implemented by~\citet{Rosindell:2010p8838}.  However, empirical evidence that fitness differences have a strong impact on species lifetimes~\citep{Adler:2010p8532} indicates that a theory combining protracted speciation with niche forces would be a much more biologically realistic solution to the species lifetime problem.  The species age problem has proven to be much more challenging to resolve within the neutral framework and may be impossible to resolve without the introduction of species asymmetries~\citep{Chisholm:2014vy}.  One such solution, based on a combination of neutral theory and Red Queen dynamics, was recently proposed by~\citet{Odwyer:2014je}.  In general, efforts to merge niche and neutral dynamics into a common framework should enhance our understanding of the impact of community dynamics on extinction~\citep{Fagan:2006p2295,Ovaskainen:2010p10809} and, eventually, may improve the accuracy of important conservation and management tools, such as the population viability analysis~\citep{Leigh:1981p11171, Lande:1993p11172, MorrisDoak:2003}.

% Our work.
In this paper, we show how niche theory and Hubbell's original formulation of neutral theory can be blended together into a single dynamical framework incorporating selection, drift (or demographic stochasticity), speciation, and dispersal -- the four principal processes of community ecology as recently underscored by~\citet{Vellend:2010ga}.  This model building exercise draws connections among many seemingly unrelated ecological population models and allows us to make quantitative predictions about the impact of niche stabilizing and destabilizing forces on population extinction times and abundance distributions.  In particular, we discuss how niche stabilization can modify species lifetime distributions by simultaneously increasing the mean values and eliminating the fat, power-law tails.  This combined effect emerges dynamically from our blended framework and makes niche stabilization a natural biological mechanism for resolving the twin problems of short species lifetimes and long species ages in the dynamics of neutral theory.  

\section*{Methods}
\label{prelim}

\subsection*{Overview}

We define a niche model based on the coupled ordinary differential equations of Lotka-Volterra dynamics, where
\begin{linenomath}
\beq
\frac{dn_i}{dt}\,=\,n_i\left(r_i-\sum_{j=1}^S a_{ij}n_j\right),
\eeq{intronichemodel}
\end{linenomath}
for each of the species, labelled $1$ through $S$.  The $n_i$ are abundances; the $r_i$ are intrinsic growth rates; and the $a_{ij}$ are per capita interaction strengths.  This article will focus on the community dynamics of ecologically similar species where all interactions are competitive, i.e.~$a_{ij}>0$ for all $i$ and $j$.  The standard requirements for coexistence emerge from the simplest case.  Consider a two-species community where intraspecific interactions are positive, i.e.~$a_{11},a_{22}>0$.  The coexisting fixed point
\begin{linenomath}
\beqa
n_1^*&=&\frac{a_{22}r_1-a_{12}r_2}{a_{11}a_{22}-a_{12}a_{21}}, \nonumber \\
n_2^*&=&\frac{a_{11}r_2-a_{21}r_1}{a_{11}a_{22}-a_{12}a_{21}},
\eeqa{intronichecoexist}
\end{linenomath}
is stable if~\citep{Vandermeer:1975p7196}
\begin{linenomath}
\beqa
r_1\frac{a_{21}}{a_{11}}&<&r_2, \nonumber \\
r_2\frac{a_{12}}{a_{22}}&<&r_1.
\eeqa{introtwostabilize}
\end{linenomath}
We will refer to these inequalities as the ``competitive stability criteria".  If all species are ecologically equivalent, such that $r_i=r$ and $a_{ij}=a$ for all $i$ and $j$, the inequalities are violated and coexistence is neutrally stable.  This phenomenology reflects Gause's axiom that species must exhibit some degree of differentiation in order to stably coexist \citep{Chase:2003}.  Furthermore, the competitive stability criteria imply that $a_{12}a_{21}<a_{11}a_{22}$, so the geometric mean of intraspecific competition must exceed the geometric mean of interspecific competition.  This requirement for niche stabilization, that species limit themselves more than other species, was emphasized by Chesson (2000) in a closely related model.  

Analysis of the niche dynamics in Eq.~\ref{intronichemodel} provides important insights on the relationship between competition and coexistence but fails to quantify the fluctuations that may lead to extinction.  Deterministic models, such as Eq.~\ref{intronichemodel}, approximate the mean dynamics of large populations where probabilities of extinction can be ignored on sufficiently short time scales.  Smaller populations require a more mechanistic approach to capture the discrete nature of birth and death events and the uncertainty surrounding the timing of those events.  In this context, exact population densities at future times cannot be predicted with certainty based on currently available information.  To be specific, given a discrete vector of known initial community abundances, $\vec{N}(t=0)=(N_1(t=0),\dots,N_S(t=0))$, we cannot predict an exact value for future abundances, $\vec{N}(t)$.  Instead, we aim to predict the probability, $P_{\vec{n}}(t)$, of finding the community in state $\vec{n}=(n_1,\dots,n_S)$ at time $t$.  $\vec{N}(t)$ is called a stochastic process and each reasonable $\vec{n}$ is referred to as an accessible state.  The time-dependent probability of each accessible state, $P_{\vec{n}}(t)$, is treated as a dynamical variable that increases or decreases based on transition rates to and from other states
\begin{linenomath}
\beqa
\frac{dP_{\vec{n}}}{dt}&=&\sum_{{\rm all~other~states}}\left({\rm Rate~of~transition~to~}\vec{n}{\rm ~from~other~state}\right) \times P_{{\rm other~state}} \nonumber \\
&&-\sum_{{\rm all~other~states}}\left({\rm Rate~of~transition~to~other~state~from~}\vec{n}\right) \times P_{\vec{n}}. 
\eeqa{}
\end{linenomath}
This equation of motion for $P_{\vec{n}}(t)$, called a ``master equation", provides a powerful platform for mechanistic modeling in ecology~\citep{Nisbet:2003}.  In particular, a master equation framework not only captures extinction dynamics due to demographic fluctuations in small populations but also yields a prescription for the mean dynamics of large populations where fluctuations are small.  Neutral theories that maintain diversity through a balance of speciation and extinction, such as Hubbell's theory, are typically formulated as master equations.  

In the sections to follow, we demonstrate that niche stabilization or destabilization can be added to neutral theories by expanding the master equations to allow for a density-dependent per capita rate of successful reproduction
\begin{linenomath}
\beq
w_{i,\vec{n}}\equiv w_{i,0}\exp\left(-\sum_{j=1}^S a_{ij}n_j/w_{i,0}\right).
\eeq{introwi}
\end{linenomath}
We will refer to $w_{i,\vec{n}}$ as the ecological fitness of species $i$, with $w_{i,0}$ being the intrinsic, density-independent, ecological fitness of species $i$ in the absence of competition.  The neutral limit only obtains when the ecological fitnesses are the same for all species.  The per capita interaction rates, $a_{ij}$, for each species $i$, are measured in the same units as the density-independent reproduction rate, $w_{i,0}$, such that the ratio $a_{ij}/w_{i,0}$ in the exponent of Eq.~\ref{introwi} is dimensionless.  On an intuitive level, this parameterization may be the simplest way to allow species to limit themselves and each other while guaranteeing that ecological fitnesses are always positive, an important requirement for the master equations to follow.  On a more formal level, we might consider $w_{i,\vec{n}}$ to be an intrinsic ecological fitness, $w_{i,0}$, multiplied by the Poisson probabilities, $\exp(-a_{ij}n_j/w_{i,0})$, that no deaths occur due to interactions with species $j$.  But regardless of its origin, we will demonstrate that the chosen form of density-dependent ecological fitness is remarkably useful in creating a non-neutral framework that combines niche dynamics and demographic stochasticity in a single theory.  Niche mechanisms may stabilize or destabilize populations on intermediate time scales, while demographic stochasticity allows for a balance of extinction and speciation over long time scales.  In the first two sections we demonstrate how density-dependent ecological fitness can stabilize or destabilize the simplest models of non-zero-sum and zero-sum stochastic dynamics.  We then expand on the zero-sum model to generate a niche-based extension of Hubbell's metacommunity and local community models.  For each model, we quantify the impact of competitive niche dynamics on mean times to extinction in metacommunities and mean times to extirpation in local communities, where immigration allows populations to recover.  After establishing this framework blending niche theory with Hubbell's original formulation of neutral theory, we discuss a niche remedy for the problems of neutral dynamics that emerges naturally from our general model.

\subsection*{A blend of niche theory and a simple birth-death process}
\label{nichebd}

The non-zero-sum dynamics of a multivariate birth-death process are governed by the master equation
\begin{linenomath}
\beq
\frac{dP_{\vec{n}}}{dt}\,=\, \sum_{i=1}^S 
\left(g_{i\vec{n}-\vec{e}_i}P_{\vec{n}-\vec{e}_i}+r_{i\vec{n}+\vec{e}_i}P_{\vec{n}+\vec{e}_i}-g_{i,\vec{n}}P_{\vec{n}}-r_{i,\vec{n}}P_{\vec{n}}\right),
\eeq{nichebdmaster}
\end{linenomath}
where $r_{i,\vec{n}}$ is the density-dependent rate of removal for an individual of species $i$, and $g_{i,\vec{n}}$ is the rate of gain.  We assume a density-dependent per capita growth rate, $w_{i,\vec{n}}$, but a density-independent per capita death rate, $d_i$, such that
\begin{linenomath}
\beqa
g_{i,\vec{n}}&\equiv& \lim_{\Delta t\rightarrow0} \frac{P\left(\vec{N}(t+\Delta t)=\vec{n}+\vec{e}_i\big|\vec{N}=\vec{n}\right)}{\Delta t}\,=\,w_{i,\vec{n}}n_i, \nonumber \\
r_{i,\vec{n}}&\equiv& \lim_{\Delta t\rightarrow0} \frac{P\left(\vec{N}(t+\Delta t)=\vec{n}-\vec{e}_i\big|\vec{N}=\vec{n}\right)}{\Delta t}\,=\,d_i n_i.
\eeqa{nichebdgr}
\end{linenomath}
This is a simple birth-death process of neutral drift if $w_{i,0}=b$, $a_{ij}=0$, and $d_i=d$ for all $i$ and $j$, such that ecological fitness is density-independent and all species are equivalent.  We now argue that density-dependent ecological fitnesses can generate stabilities or instabilities that promote or impede coexistence on intermediate time scales.

The mean dynamics of Eq.~\ref{nichebdmaster} are given by a system of ordinary differential equations where for each species $i$ (\ref{meanfieldlimit})
\begin{linenomath}
\beq
\frac{dn_i}{dt}\,=\, n_i(w_{i,\vec{n}}-d_i),
\eeq{nichebdode}
\end{linenomath}
and, in this context, $n_i=n_i(t)$ is a continuous variable.  Some familiar models are obtained in various limits.  For a single species system, a transformation to discrete time yields the Ricker model (\ref{rickerfrommeanfield}), so for two or more species, we have a generalization of the Ricker model to continuous-time community dynamics.  We also recover the  niche model upon taking a first-order expansion of the exponential that appears in $w_{i,\vec{n}}$ and identifying the $w_{i,0}-d_i$ with the $r_i$ in Eq.~\ref{intronichemodel}.  This approximation is valid where $\sum_{j=1}^Sa_{ij}n_j<<1$ for every $i$.     

The phenomenology arising from Eq.~\ref{nichebdode} is similar to results from the niche model.  In a two-species community where $a_{11},a_{22}>0$, the fixed-point abundances
\begin{linenomath}
\beqa
n_1^*&=&\frac{a_{22}w_{1,0}\log(w_{1,0}/d_1)-a_{12}w_{2,0}\log(w_{2,0}/d_2)}{a_{11}a_{22}-a_{12}a_{21}}, \nonumber \\
n_2^*&=&\frac{a_{11}w_{2,0}\log(w_{2,0}/d_2)-a_{21}w_{1,0}\log(w_{1,0}/d_1)}{a_{11}a_{22}-a_{12}a_{21}},
\eeqa{nichebdcoexist}
\end{linenomath}
are positive and stable, in the mean dynamics of Eq.~\ref{nichebdode}, for
\begin{linenomath}
\beqa
w_{2,0}\log(w_{2,0}/d_2)\frac{a_{12}}{a_{22}}<w_{1,0}\log(w_{1,0}/d_1), \nonumber \\
w_{1,0}\log(w_{1,0}/d_1)\frac{a_{21}}{a_{11}}<w_{2,0}\log(w_{2,0}/d_2).
\eeqa{nichebdstabilitycriteria}
\end{linenomath}
These criteria are identical to the competitive stability criteria if we identify $w_{1,0}\log(w_{1,0}/d_1)$ and $w_{2,0}\log(w_{2,0}/d_2)$ with $r_1$ and $r_2$, respectively.  In particular, stable coexistence hinges on familiar niche mechanisms:  species must be asymmetric such that the geometric mean of intraspecific competition exceeds the geometric mean of interspecific competition.  A niche-stabilized fixed-point in the mean dynamics of Eq.~\ref{nichebdode} corresponds to a metastability in the stochastic process governed by Eq.~\ref{nichebdmaster} that promotes coexistence on intermediate time scales and delays extinction~\citep[Ch.~8]{VanKampen:2001}.  By contrast, a niche-destabilized fixed-point in Eq.~\ref{nichebdode} corresponds to an instability in the stochastic process of Eq.~\ref{nichebdmaster} that impedes coexistence and can accelerate extinction.   

\subsection*{A blend of niche theory and a simple Moran model}
\label{nichezerosum}

Thus far, we have only considered a blend of niche theory with a non-zero-sum birth-death process.  However, Hubbell's original formulation of neutral theory is a zero-sum process that adds speciation and migration dynamics to the simple birth and death dynamics of a simple, univariate Moran model.  Before investigating a blend of niche theory and Hubbell's zero-sum neutral theory, we take the first step of integrating niche theory with a multivariate Moran model for the dynamics of $S$ species and $J$ individuals.  The accessible states, $\vec{n}=(n_1,\dots,n_S)$, must satisfy $0 \le n_i \le J$ for each $i$ and $\sum_{i=1}^Sn_i=J$.  The stochastic process follows a simple multivariate Moran model (see, e.g., \citet{Ewens:2004}), where the master equation can be written as
\begin{linenomath}
\beq
\frac{dP_{\vec{n}}}{d\tau}\,=\,\sum_{i=1}^S\sum_{j=1,j\ne i}^S\left(T_{i,j,\vec{n}+\vec{e}_i-\vec{e}_j}P_{\vec{n}+\vec{e}_i-\vec{e}_j}-T_{j,i,\vec{n}}P_{\vec{n}} \right) \Theta_{ij\vec{n}},
\eeq{nichezerosummaster}
\end{linenomath}
with a dimensionless measure of time, $\tau$.  The $\Theta_{ij\vec{n}}\equiv \Theta(J-(n_i+1))\Theta(n_j-1)$, where $\Theta(x)$ is zero for $x<0$ and one otherwise, ensure that transitions to inaccessible states are not included in Eq.~\ref{nichezerosummaster}.  A common death rate sets the overall timescale in the zero-sum transitions given by 
\begin{linenomath}
\beqa
T_{i,j,\vec{n}} &\equiv& \lim_{\Delta \tau \rightarrow0} \frac{P\left(\vec{N}(\tau+\Delta \tau)=\vec{n}-\vec{e}_i+\vec{e}_j\big|\vec{N}=\vec{n}\right)}{\Delta \tau} \nonumber \\
&=&\frac{n_i}{J}\left(\frac{w_{j,\vec{n}-\vec{e}_i}n_j}{\sum_{k=1}^Sw_{k,\vec{n}-\vec{e}_i}n_k-w_{i,\vec{n}-\vec{e}_i}}\right).
\eeqa{nichezerosumt}
\end{linenomath}
The probability of species $i$ being selected for death is just the relative abundance, $n_i/J$, of species $i$.  The probability that species $j$ recruits and establishes in the vacancy left by species $i$ is determined by drawing from the available pool of offspring in which the representation of species $k$ is determined by $w_{k,\vec{n}}n_k$.  Various subtractions in the expression for $T_{i,j,\vec{n}}$ account for the death in species $i$ that precedes the reproduction, recruitment, and establishment of species $j$.  Eq.~\ref{nichezerosummaster}, with Eq.~\ref{nichezerosumt}, reduces to a simple model of zero-sum neutral drift in a symmetric community where $w_{i,0}=b$ and $a_{ij}=a_j$ for all $i$ and $j$.  The former condition ensures that intrinsic fitnesses are equivalent, while the latter ensures the absence of a niche stabilizing mechanism; together, these conditions are the requirements for neutrality highlighted by \citet{Adler:2007p8512}.  Breaking these symmetries allows for niche stabilization and destabilization.

The mean dynamics of Eq.~\ref{nichezerosummaster}, when written in terms of $p_i\equiv n_i/J$, take the form of a Levins model~\citep{Levins:1971p7231}.  For sufficiently weak competitive asymmetries such that $w_{i,\vec{n}}<<\sum_{k=1}^S w_{k,\vec{n}}n_k$ and $a_{ij}<<\sum_{k=1}^S a_{ik}n_k$ for every $i$ and $j$, we have (\ref{meanfieldlimit}) 
\begin{linenomath}
\beq
\frac{dp_i}{d\tau}\,=\, c_i(\vec{p}) p_i \left(1-p_i\right) -\sum_{j=1,j\ne i}^Sc_{j}(\vec{p})p_ip_j,
\eeq{nichezerosumode}
\end{linenomath}
where
\begin{linenomath}
\beq
c_{i}(\vec{p})\,=\,\frac{w_{i}(\vec{p})}{J\sum_{k=1}^Sw_{k}(\vec{p}) p_k}.
\eeq{nichezerosumdefs}
\end{linenomath}
Each $w_{i}(\vec{p})$ is obtained from the expression for $w_{i\vec{n}}$ with the substitution of $p_i J$ for $n_i$.  The mean dynamics of Eq.~\ref{nichezerosumode} prescribe zero-sum niche dynamics for a community with any given number of species and individuals (\ref{meanfieldzerosumproof}).  The phenomenology is remarkably similar to the niche model for a two-species system, and we will refer to Eq.~\ref{nichezerosumode} as a ``Levins niche model".  Fixed-point relative abundances  
\begin{linenomath}
\beqa
p_1^*&=&\frac{B_2}{B_1+B_2}, \nonumber \\
p_2^*&=&1-p_1^*,
\eeqa{metadeterfp}
\end{linenomath}
where
\begin{linenomath}
\beqa
B_1&=&a_{11}J/w_{1,0}-a_{21}J/w_{2,0}-\log(w_{1,0}/w_{2,0}), \nonumber \\
B_2&=&a_{22}J/w_{2,0}-a_{12}J/w_{1,0}+\log(w_{1,0}/w_{2,0}),
\eeqa{metadeterbdefs}
\end{linenomath}
are positive and stable for 
\begin{linenomath}
\beqa
w_{1,0}\frac{e^{a_{21}J/w_{2,0}}}{e^{a_{11}J/w_{1,0}}}&<&w_{2,0}, \nonumber \\
w_{2,0}\frac{e^{a_{12}J/w_{1,0}}}{e^{a_{22}J/w_{2,0}}}&<&w_{1,0},
\eeqa{metadetervandermeer}
\end{linenomath}
which implies that $a_{12}a_{21}<a_{11}a_{22}$.  The correspondence is clear to the competitive stability criteria and to the familiar mechanisms of niche stabilization and destabilization.  We find that zero-sum assumptions do not have a strong impact on niche dynamics for large communities where the mean dynamics are a good approximation to the underlying stochastic processes of birth, death, and competition.  This result parallels the conclusion of \citet{Etienne:2007p6358} that zero-sum assumptions do not have a strong impact on neutral dynamics in high-diversity communities.  

\section*{Results}

The preceding section has demonstrated how a standard niche model can be incorporated into standard models for demographic stochasticity.  We now apply that formalism to quantify the impacts of competition on extinction and to blend a standard niche model with Hubbell's original formulation of neutral theory.

\subsection*{Impacts of competition on extinction}
\label{question}

If the first species competes with one or more symmetric species, the full multivariate stochastic dynamics of Eq.~\ref{nichezerosummaster} can be reduced to a univariate master equation for the marginal dynamics of the first species (\ref{numericalintegration}).  Fig.~\ref{moranstochintegration} plots the temporal evolution of the conditional abundance probability distribution, $P_{cn_1,\tau}$, as determined by Eq.~\ref{asymmPnM} after excluding situations in which the first species reaches total extinction or  complete dominance.  Panels 1a, 1b, and 1c correspond, respectively, to scenarios of neutrality, interspecific exceeding intraspecific competition, and intraspecific exceeding interspecific competition.  In all cases, the conditional probabilities approach a quasi-stationary distribution at long times.  Compared to the neutral case, an excess of interspecific competition destabilizes the system by inducing a bimodal distribution where one of the two species dominates over short time scales with a high probability.  This bistability corresponds to an unstable fixed-point of the mean dynamics and signals a stochastic Allee effect that accelerates the extinction of rare species.  By contrast, an excess of intraspecific competition generates a single peak in the conditional distribution.  This peak is a metastability that corresponds to a stable fixed-point of the mean dynamics.  In the $J\rightarrow\infty$ limit, the peaks of $P_{cn_1,\tau}$ approach delta functions, demographic stochasticity vanishes, and the mean dynamics of the Levins niche model (Eq.~\ref{nichezerosumode}) are recovered.  For finite $J$, absorbing states guarantee monodominance by a single species at long times, while niche dynamics accelerate or delay extinction on intermediate time scales.  

In Online Resource 5, we extend the multivariate Moran model and its mean dynamics to allow for empty space.  Summations run to $S+1$ and the stochastic variable $N_{S+1}(\tau)$ tracks the number of unoccupied patches.  For large zero-sum communities where the vast majority of spaces are unoccupied, population dynamics are well-approximated, after a rescaling of time, by non-zero-sum dynamics.  Fig.~\ref{theorysumm} summarizes connections among the blended stochastic processes and deterministic models developed in the first part of this paper.  All models contain a neutral theory in the symmetric limit.

Of course, neither of the two stochastic frameworks discussed thus far offers a valid model of biodiversity over long time scales.  In the simple birth-death process, each species will either become extinct or approach infinite abundance.  In the simple Moran model, a single species will dominate.  The next two sections extend the Moran model of Eq.~\ref{nichezerosummaster} by allowing speciation or migration to remove the absorbing states at monodominance that inhibit the long-term maintenance of species diversity.  These non-neutral frameworks inherit the niche dynamics of our simple Moran model but yield Hubbell's theory of metacommunities and local communities in the neutral limit.

\subsection*{A blend of niche theory and Hubbell's neutral theory of the metacommunity}
\label{meta}

A previous extension of Hubbell's metacommunity theory, with point-speciation, allowed for asymmetries in ecological fitness and speciation probability (Box 1, Eq.~\ref{basymmPnvecM}).  We now expand on that asymmetric framework by introducing density-dependence in ecological fitness.  To do this, we replace all the density-independent $w_i$ that appear in Eq.~\ref{basymmPnvecM} by the density-dependent $w_{i,\vec{n}-\vec{e}_i}$ to obtain a new master equation.  The simple Moran model is recovered when all speciation probabilities vanish, and a multivariate formulation of Hubbell's metacommunity theory is included as the symmetric limit.  The mean dynamics approximating our non-neutral metacommunity dynamics can be written as a Levins model, and we once again assume weak competitive asymmetries to obtain
\begin{linenomath}
\beq
\frac{dp_i}{d\tau}\,=\, s_i(\vec{p}) +c_i(\vec{p}) p_i \left(1-p_i\right)- e(\vec{p}) p_i -\sum_{j=1,j\ne i}^Sc_{j}(\vec{p})p_ip_j +{\cal O}\left(\frac{1}{S}\right),
\eeq{metagenode}
\end{linenomath}
where ${\cal O}\left(1/S\right)$ indicates additional terms that are negligible as the number of species becomes large and
\begin{linenomath}
\beqa
s_i(\vec{p})&=&\frac{\nu_i }{J_M}\frac{w_i(\vec{p})}{\sum_{k=1}^Sw_k(\vec{p})}, \nonumber \\
c_i(\vec{p})&=&\frac{1-\nu_i}{J_M}\frac{w_i(\vec{p})}{\sum_{k=1}^Sw_k(\vec{p}) p_k}, \nonumber \\
e(\vec{p})&=&\sum_{j=1}^S\frac{\nu_j}{J_M}\frac{w_j(\vec{p})}{\sum_{k=1}^Sw_k(\vec{p})}.
\eeqa{metagendefs}
\end{linenomath}
In addition to colonization and competition, we now find a source term, with coefficient $s_i(\vec{p})$, and an extinction term, with coefficient $e(\vec{p})$, arising from nonzero speciation probabilities.  Eq.~\ref{metagenode} is similar to the replicator-mutator equation in population genetics~\citep{Ewens:2004}:  all possible species must be enumerated, and in principle, an individual of any given species can give birth to an individual of any other species, consistent with the assumption of weak competitive asymmetries.

The mean dynamics of Eq.~\ref{metagenode} approximate non-neutral metacommunity dynamics given any fixed number of species and individuals.  In the neutral limit, we find
\begin{linenomath}
\beq
\frac{dp_i}{d\tau}\,=\, \frac{\nu}{J_M}\left(\frac{1}{S}-p_i\right),
\eeq{metagenodeneutral}
\end{linenomath}
such that all species completely decouple.  In the limit of an infinite number of species and individuals ($S, J_M\rightarrow \infty$), the $p_i$ remain fixed at their initial values, which is the deterministic limit of neutral drift.  For a nearly neutral metacommunity with low levels of speciation, a fixed-point analysis of Eq.~\ref{metagenode} yields small corrections to the stability criteria of the simple Moran model (\ref{correctionstostabilizationcriteria}), so we expect that our non-neutral metacommunity model inherits niche stabilization and destabilization mechanisms.

Confirmation is provided by a calculation of extinction times for the asymmetric species in a nearly neutral metacommunity (\ref{extinctiontimes}).  Fig.~\ref{metastochdeter}a plots $\tau_E$ against $p_{1,0} \equiv n_{1,0}/J_M$, while Fig.~\ref{metastochdeter}b plots the corresponding flows, $dp_1/d\tau$ versus $p_1$, of the mean dynamics.  In the fully neutral limit, demographic stochasticity drives all species toward extinction in the absence of a stabilizing mechanism~\citep{Chesson:2000p5531,PHubbell:2001p4284}, and the mean dynamics yield a single stable fixed point at $p_1=0$.  For the case where interspecific exceeds intraspecific competition, $\tau_E$ falls below (above) neutral expectations when $p_{1,0}$ is low (high).  The existence of an inflection point in the $\tau_E$ versus $p_{1,0}$ curve signals a stochastic Allee effect, as discussed by~\citet{Dennis:1989,Dennis:2002p8840} and \citet{Allen:2005p8841}.  Indeed, the mean dynamics exhibit upper and lower stable fixed points separated by an unstable threshold.  The underlying bistability of the stochastic theory allows large initial populations of the asymmetric species to remain dominant on intermediate time scales, but large fluctuations below the unstable threshold rapidly reduce mean times to extinction.  Only intraspecific exceeding interspecific competition raises $\tau_E$ above neutral expectations for all values of $p_{1,0}$.  The corresponding mean dynamics include a stable fixed point that promotes stable coexistence over intermediate time scales as well as an unstable fixed point that repels drift toward extinction. 

\subsection*{A blend of niche theory and Hubbell's neutral theory of the local community}
\label{local}

We now expand on an asymmetric extension of Hubbell's local community theory, as described in Box 1, by allowing for density-dependence in ecological fitness.  The master equation is given by Eq.~\ref{basymmPnvecL} with the replacement of each $w_i$ by $w_{i,\vec{n}-\vec{e}_i}$.  The simple Moran model is recovered in the limit where all immigration probabilities vanish, and the multivariate formulation of Hubbell's local community theory \citep[p.~128]{PHubbell:2001p4284} is included as the symmetric limit.  The mean dynamics approximating our non-neutral local community dynamics can be written as a Levins model, and we assume weak competitive asymmetries to obtain
\begin{linenomath}
\beq
\frac{dp_i}{d\tau}\,=\, s_i(\vec{p}) +c_i(\vec{p}) p_i \left(1-p_i\right)- e(\vec{p}) p_i -\sum_{j=1,j\ne i}^Sc_{j}(\vec{p})p_ip_j,
\eeq{localgenode}
\end{linenomath}
where
\begin{linenomath}
\beqa
s_i(\vec{p})&=&\frac{m_i }{J_L}\frac{w_i(\vec{p}) x_i}{\sum_{k=1}^Sw_k(\vec{p})x_k}, \nonumber \\
c_i(\vec{p})&=&\frac{1-m_i}{J_L}\frac{w_i(\vec{p})}{\sum_{k=1}^Sw_k(\vec{p}) p_k}, \nonumber \\
e(\vec{p})&=&\sum_{j=1}^S\frac{m_j}{J_L}\frac{w_j(\vec{p}) x_j}{\sum_{k=1}^Sw_k(\vec{p}) x_k}.
\eeqa{localgendefs}
\end{linenomath}
The $x_i$ are relative metacommunity abundances that are assumed to be fixed in the dynamics of the local community (see Eq.~\ref{basymmPnvecL}).  In Eq.~\ref{localgenode}, the source and extinction terms arise from nonzero immigration probabilities, and the mean dynamics approximate non-neutral local community dynamics given any fixed number of species and individuals.  In the neutral limit, we have~\citep{Vallade:2003p3851}
\begin{linenomath}
\beq
\frac{dp_i}{d\tau}\,=\, \frac{m}{J_L}(x_i-p_i),
\eeq{localgenodeneutral}
\end{linenomath}
such that all species completely decouple and local relative community abundances track relative metacommunity abundances at equilibrium.  Similarly, in the stochastic formulation of Hubbell's local community, the expected relative abundance of species $i$ is $x_i$ at equilibrium~\citep{PHubbell:2001p4284}.  This provides a built-in mechanism for stabilizing coexistence:  given a nonzero probability of immigration, high levels of diversity in the metacommunity guarantee high levels in the local community when abundances are averaged over long time scales.  However, low immigration probabilities generate large fluctuations that destabilize local diversity on shorter time scales and prolong periods of extirpation.  Expanding the equilibrium expression for relative fluctuation amplitude in Hubbell's theory~\citep{Vallade:2003p3851} (see the two equations that follow their Eq.~7 and precede their Sec.~III), we find (\ref{localcv})
\begin{linenomath}
\beq
\frac{\sqrt{\langle N_i - \langle N_i \rangle \rangle^2}}{\langle N_i \rangle}\,=\,\frac{1}{\sqrt{J_L}}\sqrt{\frac{1-x_i}{mx_i}}+{\cal O}(J_L^{-3/2}),
\eeq{localdetercv}
\end{linenomath}
so, for sufficiently small values of $m$, population fluctuations equal to the total number of individuals in the community become common, which implies a bimodal stationary distribution with peaks at extirpation and monodominance.  

Our non-neutral framework extends Hubbell's theory to allow for a niche mechanism that, despite low levels of immigration, promotes coexistence and lengthens the time to extirpation.  To demonstrate this, we calculate a stationary distribution, $P_{n_1}^*$, for the asymmetric species in a nearly neutral local community (\ref{localnnstationary}).  Averaging over long-term stochastic fluctuations, the relative amount of time that the asymmetric species spends in state $n_1$ is equal to $P_{n_1}^*$.  Given $m_1=m_2<<1$, Fig.~\ref{localstochdeter}a plots stationary distributions for the asymmetric species in three scenarios, while Fig.~\ref{localstochdeter}b plots corresponding flows in the mean dynamics.  In the fully neutral limit, the stationary distribution is bimodal with peaks at extirpation and monodominance, and low probabilities of coexistence.  When interspecific exceeds intraspecific competition, the mean dynamics exhibit an Allee effect where the unstable threshold at intermediate abundance corresponds to a further reduction in probabilities of coexistence for the stationary distribution~\citep{Dennis:1989}.  By contrast, intraspecific exceeding interspecific competition induces a tri-stable distribution with a metastability for coexistence that reduces periods of extirpation and monodominance.  That metastability corresponds to a stable fixed point of the mean dynamics, and for low levels of immigration, stability criteria are well-approximated by results for the simple Moran model (\ref{correctionstostabilizationcriteria}).   

\subsection*{Resolving species lifetime and age problems in the dynamics of neutral theory}
\label{applifeage}

% Pigolotti
Working from a preprint of this paper, \citet{Pigolotti:2013gk} have studied our blend of niche theory and Hubbell's metacommunity theory in the limit of totally symmetric competition with deviations from the neutral limit governed by a single parameter.  Species abundance distributions can be calculated analytically in this limit and the species lifetime distribution can be studied numerically.  When intraspecific competition exceeds interspecific competition,~\citet{Pigolotti:2013gk} demonstrate that mean species lifetimes increase above neutral expectations and the variance of species lifetimes decreases.  Importantly, niche stabilization truncates the fat, power-law tail of the species lifetime distribution, as predicted by neutral theory, with an exponential cutoff.  In this way, niche stabilization provides a natural biological mechanism for resolving, simultaneously, the species lifetime and species age problems in the dynamics of neutral theory.  

%Ricklefs
Other limits and extensions of our general theory may provide additional solutions to the species lifetime and age problems in neutral theory that are both analytically tractable and biologically well-motivated.  For example, \citet{Ricklefs:2003p6692} suggested that the lifetime problem could be resolved if rare species enjoy a competitive advantage that increases persistence and that the age problem could be resolved if abundant species suffer occasional competitive disadvantage from the speciation of superior competitors.  We have shown that the former effect emerges from our blended theory when intraspecific exceeds interspecific competition, while the latter effect emerges when interspecific exceeds intraspecific competition.  Adding temporal environmental variation to our theory might allow the suggestion of \citet{Ricklefs:2003p6692} to be implemented in a quantitative framework.  Indeed, there is mounting evidence that temporal environmental variation should not be ignored in building biologically realistic models of community dynamics~\citep{Chisholm:2014jv, Kalyuzhny:2014tu}.

%Further work remains to enable rigorous, quantitative tests for this stabilizing mechanism in field data.  For example, 

\section*{Discussion}
\label{discuss}

In his influential review of mechanisms for the maintenance of species diversity, \citet{Chesson:2000p5531} notes that, ``Models of unstable coexistence, in which species diversity slowly decays over time, have focused almost exclusively on equalizing mechanisms. These models would be more robust if they also included stabilizing mechanisms, which arise in many and varied ways but need not be adequate for full stability of a system. Models of unstable coexistence invite a broader view of diversity maintenance incorporating species turnover".  With the simple introduction of asymmetries in density-dependent ecological fitnesses, our blend of niche theory and Hubbell's neutral theory  embeds both stabilizing and destabilizing niche mechanisms within a neutral model of unstable drift, speciation, and dispersal~\citep{Vellend:2010ga}.  We have discussed how niche stabilizing forces can resolve, simultaneously, two major problems in the dynamics of neutral theory, namely predictions of species lifetimes that are too short and species ages that are too long.

This niche remedy is a natural and biologically well-motivated solution.  After all, much of the controversy surrounding Hubbell's theory centers on the assumption of per capita equivalence in ecological fitness (see, e.g.,~\citet{Leibold:2006p6367}).  \citet{PHubbell:2001p4284} argued that life-history tradeoffs constrain niche differentiation to fitness-invariant manifolds.  In our non-neutral framework, Hubbell's theory lies along hyperplanes of parameter space where $w_{i,0}=w$ and $a_{ij}=a_j$ for all $i$ and $j$.  In this context, larger questions about the evolution of neutral or nearly neutral community dynamics boil down to a search for mechanisms driving the convergence of ecological fitnesses and interaction strengths.  Toward this end, future work should integrate mechanistic niche models, such as the consumer-resource models of~\citet{Tilman:1980p9016,Tilman:1982p7101}, into a stochastic theory of unstable coexistence.  

Many additional directions in model-building remain to be explored.  In order to obtain testable predictions based on observations of non-cryptic, recognizable species, a protracted speciation mechanism, similar to~\citet{Rosindell:2010p8838}, could be added to our general framework.  Within this expanded framework, the techniques of~\citet{Chisholm:2014vy} might allow for analytical calculations of expected species ages as a function of species abundance.  Furthermore, because our nested analytical framework provides a natural interpolation between niche and neutral dynamics, it could serve as a pivot point from which additional features, already well-studied in niche-based models, could be incorporated into neutral models of speciation and drift. For example, niche-based models similar to Eq.~\ref{nichezerosumode} have been modified to include effects of habitat destruction~\citep{Tilman:1994p7531,Tilman:1997wm}, leading to the discovery of the ``extinction debt" in which species do not go extinct until long after habitat destruction takes place.  This critical issue remains unexplored in the framework of neutral theory, where extinction debts could prolong species persistence times beyond those expected from the effects of demographic stochasticity alone. 

Other work, such as~\citet{Melbourne:2008p9904}, highlights the need for model-building to move beyond considerations of simple demographic noise to incorporate explicit, parametric models of environmental stochasticity and demographic heterogeneity, among other sources of variation.  \cite{ODwyer:2009p4394} have integrated size-structure into Hubbell's theory, and our non-neutral model of unstable coexistence might be expanded in similar ways.  Further efforts should also be made to incorporate niche stabilization into spatially explicit models of unstable coexistence.  Incorporating space into Hubbell's theory has generated analytical predictions for clustering \citep{Houchmandzadeh:2003p7113}, beta-diversity \citep{Chave:2002p10147,Zillio:2005p10150}, and a tri-phasic species-area relationship \citep{ODwyer:2009p6371}.  The model-building efforts of~\citet{Economo:2008p4393}, \citet{Muneepeerakul:2008p2327},~\citet{Babak:2009p7109}, and~\citet{Vanpeteghem:2010p8333} implement neutral dynamics over a distributed network of patches.  This work should be expanded to incorporate asymmetries and mechanisms of niche stabilization and destabilization.  Advances in stochastic modeling, both spatially-explicit and spatially-implicit, will underscore the importance of extended transients in ecological dynamics~\citep{Hastings:2004p1529, Hastings:2010p12514} and provide novel insights on the early warning signals of extinction in deteriorating environments~\citep{Drake:2010p11600}.

The study of neutral theories has led to suggestions that emergent regularities, such as unimodal RSA distributions, imply a simplicity in the underlying dynamics of ecological communities.  But increasingly, the opposite perspective is receiving attention.  How much complexity can ecological theory embrace at the level of individuals without losing the ability to make testable predictions about large-scale patterns of biodiversity?  Can we rigorously demonstrate how large-scale biogeographic regularities can emerge from the many irregularities of individual-based ecological interactions?  Significant progress will be made if novel analysis techniques can be developed for the macroscopic scaling of multi-species stochastic processes.

%\section*{Box 1:  Limit cycles}
%\label{predprey}
%
%If we consider Eq.~\ref{nichebdode} as a prescription for two-species predator-prey dynamics, a stable, self-sustaining limit cycle may arise, as shown in Fig.~\ref{limitcycle}.  Although stable limit cycles can be found in two-species systems with a functional response (see, e.g., \citet{Kot:2001}), they do not occur in the two-species niche model \citep{HASTINGS:1978p10096}, which, as noted above, approximates the dynamics of Eq.~\ref{nichebdode} when abundances are sufficiently small.   In fact, our numerical work indicates that the stable limit cycle of Fig.~\ref{limitcycle} vanishes when the exponential of density-dependent ecological fitness is approximated by a truncated power series of any order.  

\makeatletter
\@addtoreset{equation}{section}
\makeatother
\renewcommand{\theequation}{B1.\arabic{equation}}

%Latex needs a numbered section to reset the equation numbers.  Here, I initialize a dummy section, but I use white text, so the lone ``1" doesn't appear on the pdf.
{\color{White}{\section{}}}

\section*{Box 1:  An asymmetric extension of Hubbell's neutral theory}
\label{asymm}

Previous work, in collaboration with Nico Temme and Tim Keitt, introduced an analytical model of asymmetric, zero-sum community dynamics that contains Hubbell's neutral theory of biodiversity in the symmetric limit~\citep{Noble:2011p10042}.  Our asymmetric metacommunity ($M$) retains Hubbell's point-speciation model but allows for variation in ecological fitnesses, the $w_i$, and speciation probabilities, the $\nu_i$, across species.  Dynamics are governed by the master equation
\begin{linenomath}
\beq
\frac{dP^M_{\vec{n}}}{d\tau}\,=\,\sum_{i=1}^{S}\sum_{j=1,j\ne i}^{S}\left(T^M_{ij\vec{n}+\vec{e}_i-\vec{e}_j}P^M_{\vec{n}+\vec{e}_i-\vec{e}_j}-T^M_{ji\vec{n}}P^M_{\vec{n}} \right) \Theta^M_{ij},
\eeq{basymmPnvecM}
\end{linenomath}
where $S$ is the total number of possible species, $\tau$ is a dimensionless measure of time, the $\vec{e}_i$ are $S$-dimensional unit vectors, and the
\begin{linenomath}
\beq
T^M_{ij\vec{n}} \,=\, \frac{n_i}{J_M}\left((1-\nu_j)\frac{w_jn_j}{\sum_{k=1}^{S}w_kn_k-w_i}+\nu_j\frac{w_j}{\sum_{k=1}^{S}w_k}+{\cal O}\left(\frac{1}{S}\right)\right),
\eeq{basymmtM}
\end{linenomath}
are one-step transition probabilities for the removal of species $i$ followed by the addition of species $j$, where ${\cal O}\left(1/S\right)$ indicates the presence of additional terms that are negligible as the number of species becomes large.  The number of individuals, $J_M$, is fixed, and the accessible states are abundance vectors, $(n_1,\dots,n_S)$, where $\sum_{i=1}^S n_i=J_M$ and $0\leq n_i \leq J_M$.  The $\Theta^M_{ij}=\Theta(J_M-(n_i+1))\Theta(n_j-1)$, where $\Theta(x)$ is zero for $x<0$ and one otherwise, ensure that transitions to inaccessible states are not included in Eq.~\ref{basymmPnvecM}.  Note the absence of absorbing states:  species have nonzero probabilities of re-introduction following an extinction event.  However, as the number of possible species becomes appropriately large, the probability of re-introduction becomes vanishingly small.  

Our asymmetric local community ($L$) allows for variation in ecological fitnesses, the $w_i$, and immigration probabilities, the $m_i$, given fixed relative metacommunity abundances, the $x_i$.  Dynamics are governed by the multivariate master equation
\begin{linenomath}
\beq
\frac{dP_{\vec{n}}^L}{d\tau}\,=\,\sum_{i=1}^S\sum_{j=1,j\ne i}^S\left(T_{i,j,\vec{n}+\vec{e}_i-\vec{e}_j}^LP_{\vec{n}+\vec{e}_i-\vec{e}_j}^L-T_{j,i,\vec{n}}^LP_{\vec{n}^L}^L \right) \Theta^L_{ij},
\eeq{basymmPnvecL}
\end{linenomath}
where 
\begin{linenomath}
\beq
T_{i,j,\vec{n}}^L \,=\, \frac{n_i}{J_L}\left((1-m_j)\frac{w_jn_j}{\sum_{k=1}^Sw_kn_k-w_i}+m_j\frac{w_jx_j}{\sum_{k=1}^Sw_kx_k}\right).
\eeq{basymmtL}
\end{linenomath}
The number of individuals, $J_L$, is fixed, and the accessible states are abundance vectors, $(n_1,\dots,n_S)$, where $\sum_{i=1}^S n_i=J_L$ and $0\leq n_i \leq J_L$.  The $\Theta^L_{ij}=\Theta(J_L-(n_i+1))\Theta(n_j-1)$ ensure that transitions to inaccessible states are not included in Eq.~\ref{basymmPnvecL}.  Nonzero values for the $m_i$ guarantee the absence of absorbing states and promote coexistence via mass-effects.  In Hubbell's neutral theory, at equilibrium, expected relative abundances in the local community equal relative abundances in the metacommunity.  By introducing interspecific heterogeneity in ecological fitness and migration probability, our asymmetric theory allows the expected local community composition to differ from the metacommunity composition.  

\section*{Acknowledgments}
\label{ack}
This work originated in the lab of W.~F.~F.  Comments from Rampal Etienne, Alan Hastings, Marcel Holyoak, and Tim Keitt as well as a thorough and insightful review from Ryan Chisholm significantly improved the quality of this manuscript.  A.~E.~N.~benefited from productive conversations with Sivan Leviyang and correspondence with Juliana Berbert.  Our work was partially supported by the James S. McDonnell Foundation through their Studying Complex Systems grant (220020138) to W.~F.~F.  

\makeatletter
\@addtoreset{equation}{subsection}
\makeatother
\renewcommand{\theequation}{OR.\arabic{subsection}.\arabic{equation}}

\section*{Online Resources}

\makeatother
\renewcommand{\thesubsection}{Online Resource \arabic{subsection}}

\subsection{Mean dynamics}
\label{meanfieldlimit}

\noindent According to Eq.~3.5.14 of \citet{Gardiner:2004}, the mean dynamics of a multivariate Markov process is a system of ordinary differential equations given by an approximation that ignores correlations
\begin{linenomath}
\beq
\frac{dn_i}{dt}\,\sim\,A_i,
\eeq{simflode}
\end{linenomath}
for each species $i$. The 
\begin{linenomath}
\beq
n_i\equiv\sum_m m P\left(\vec{N}_i=m\right)
\eeq{} 
\end{linenomath}
are promoted from a discrete index to a continuous variable and
\begin{linenomath}
\beq
A_i+{\cal O}(\epsilon)\,=\,\lim_{\Delta t \rightarrow 0} \frac{1}{\Delta t}\int_{|\vec{m}-\vec{n}|<\epsilon}d\vec{m}~(m_i-n_i)P\left(\vec{N}(t+\Delta t)=\vec{m}\big|\vec{N}(t)=\vec{n}\right),
\eeq{simflai}
\end{linenomath}
are the first-order jump moments as defined by Eq.~3.4.2 of \citet{Gardiner:2004}.  For the transition probabilities of Eq.~\ref{nichebdgr}, we find
\begin{linenomath}
\beq
P\left(\vec{N}(t+\Delta t)=\vec{m}\big|\vec{N}=\vec{n}(t)\right)\,=\,\sum_{j=1}^S\big(\delta((\vec{m}+\vec{e}_j)-\vec{n})g_{j,\vec{n}}+\delta((\vec{m}-\vec{e}_j)-\vec{n})r_{j,\vec{n}}\big)\Delta t+o(\Delta t),
\eeq{simflpbd}
\end{linenomath}
so
\begin{linenomath}
\beq
A_i\,=\,g_{i,\vec{n}}-r_{i,\vec{n}},
\eeq{simflabd}
\end{linenomath}
and Eq.~\ref{simflode} yields Eq.~\ref{nichebdode}.  For the transition probabilities of Eq.~\ref{nichezerosumt}, we find
\begin{linenomath}
\beqa
P\left(\vec{N}(\tau+\Delta \tau)=\vec{m}\big|\vec{N}=\vec{n}(\tau)\right)=\sum_{k=1}^S\sum_{j=1,j\ne k}^S&&\hskip-0.6cm \big(\delta((\vec{m}-\vec{e}_j+\vec{e}_k)-\vec{n})T_{j,k,\vec{n}}\big. \nonumber \\
&&\hskip-0.6cm \big.+\delta((\vec{m}-\vec{e}_k+\vec{e}_j)-\vec{n})T_{k,j,\vec{n}}\big)\Delta \tau+o(\Delta \tau), \nonumber \\
\eeqa{simflmoran}
\end{linenomath}
so
\begin{linenomath}
\beq
A_i\,=\,\sum_{j=1,j\ne i}\left(T_{j,i,\vec{n}}-T_{i,j,\vec{n}}\right),
\eeq{simflamoran}
\end{linenomath}
and Eq.~\ref{simflode}, with $t\rightarrow \tau$, yields Eq.~\ref{nichezerosumode}, given $p_i \equiv n_i/J$ and the assumption of sufficiently weak competitive asymmetry such that $w_{i,\vec{n}}<<\sum_{k=1}^S w_{k,\vec{n}}n_k$ and $a_{ij}<<\sum_{k=1}^S a_{ik}n_k$ for every $i$ and $j$.  A Kramers-Moyal expansion or Van Kampen system size expansion yields mean dynamics identical to the ones derived here (see, e.g., \citet[p.~251]{Gardiner:2004}).

\subsection{Obtaining the Ricker model from the mean dynamics of a simple birth-death process}
\label{rickerfrommeanfield}

\noindent  Eq.~\ref{nichebdode} prescribes the single-species dynamics
\begin{linenomath}
\beq
\frac{dn_1}{dt}\,=\,n_1(w_{1,0} e^{-a_{11}n_1/w_{1,0}}-d_1).
\eeq{sirfmf}
\end{linenomath}
Let $\tau=d_1t$ and descritize the derivative to obtain
\begin{linenomath}
\beqa
n_{1\tau+1}&=&n_{1\tau}+n_{1\tau}\left(\frac{w_{1,0}}{d_1} e^{-a_{11}n_{1\tau}/w_{1,0}}-1\right), \nonumber \\
&=& n_{1\tau}e^{r(1-n_{1\tau}/K)},
\eeqa{sirfmfdiscrete}
\end{linenomath}
where
\begin{linenomath}
\beqa
r&=& \log (w_{1,0}/d_1), \nonumber \\
K&=&\frac{w_{1,0}}{a_{11}}\log (w_{1,0}/d_1).
\eeqa{sirfmfdef}
\end{linenomath}
Eq.~\ref{sirfmfdiscrete} is the Ricker model.

\subsection{The mean dynamics of a Moran model retains the zero-sum rule}
\label{meanfieldzerosumproof}

\noindent Summing Eq.~\ref{nichezerosumode} over all species, we obtain
\begin{linenomath}
\beqa
\sum_{i=1}^S\frac{dp_i}{d\tau}&=&\sum_{i=1}^S c_ip_i\left(1 - \sum_{j=1}^Sp_j \right).
\eeqa{simfzs_moran}
\end{linenomath}
If $\sum_{j=1}^S p_i(0)=1$, then $\sum_{i=1}^Sdp_i/d\tau\big\vert_{\tau=0}=0$, which is sufficient to guarantee that $\sum_{j=1}^S p_i(\tau)=1$ for all $\tau$.

\subsection{Dynamics of a simple Moran model}
\label{numericalintegration}

\noindent In the $S=2$ case of Eq.~\ref{nichezerosummaster}, the stochastic dynamics can be written, without approximation, as a univariate master equation for the marginal distribution of the first species
\begin{linenomath}
\beqa
\frac{P_{n_1}}{d\tau}&=&g_{n_1-1}\Theta(n_1-1)P_{n_1-1} + r_{n_1+1}\Theta(J- (n_1+1))P_{n_1+1} \nonumber \\
&& - \left(g_{n_1}\Theta(J-(n_1+1))+r_{n_1}\Theta(n_1-1)\right)P_{n_1},
\eeqa{asymmPnM}
\end{linenomath}
with
\begin{linenomath}
\beqa
g_{n_1}&\equiv&T_{2,1,(n_1,n_2)} \nonumber \\
&=&\frac{J-n_1}{J}\left(\frac{e^{-((B_1+B_2)n_1/J-B_2-a_{12}/w_{1,0}+a_{22}/w_{2,0})} n_1}{e^{-((B_1+B_2)n_1/J-B_2-a_{12}/w_{1,0}+a_{22}/w_{2,0})} n_1+ J-n_1-1}\right), \nonumber \\
r_{n_1} &\equiv& T_{1,2,(n_1,n_2)} \nonumber \\
&=&\frac{n_1}{J}\left(\frac{J-n_1}{e^{-((B_1+B_2)n_1/J-B_2+a_{21}/w_{2,0}-a_{11}/w_{1,0})} (n_1-1)+J-n_1}\right). \nonumber \\
\eeqa{asymmgrM}
\end{linenomath}
This master equation also governs marginal dynamics for the asymmetric species in a nearly neutral community where all other species, labelled $2$ thru $S$, are symmetric (see \citet{Noble:2011p10042}).  
\\
\\
\noindent To calculate the temporal evolution of {\it conditional} abundance probability distributions, as plotted in Fig.~\ref{moranstochintegration}, we start by discretizing the univariate birth-death process of Eq.~\ref{asymmPnM} to obtain
\begin{linenomath}
\beqa
P_{n_1,\tau+1}&=&g_{n_1-1}\Theta(n_1-1)P_{n_1-1,\tau} + r_{n_1+1}\Theta(J- (n_1+1))P_{n_1+1,\tau} \nonumber \\
&& + \left(1-g_{n_1}\Theta(J-(n_1+1))-r_{n_1}\Theta(n_1-1)\right)P_{n_1,\tau} \nonumber \\
&=&\sum_{m=0}^{J}P_{m,\tau}W_{mn_1},
\eeqa{sinumintdiscrete}
\end{linenomath}
where
\begin{linenomath}
\beq
W\,=\,\left(\begin{array}{ccccccc}
1-g_0& g_0& 0& \cdots &0&0&0\\
r_1& 1-r_1-g_1& g_1&\cdots &0&0&0\\
\vdots& \vdots& \vdots& \ddots &\vdots &\vdots&\vdots\\
0& 0& 0& \cdots& r_{J-1}& 1-r_{J-1}-g_{J-1}& g_{J-1}\\
0& 0& 0& \cdots& 0& r_{J}& 1-r_{J} 
\end{array}\right).
\eeq{sinumintw}
\end{linenomath}
The unconditioned abundance probability distribution at any integer time $\tau$, denoted $\vec{P}_{\tau}$, is given by
\begin{linenomath}
\beq
\vec{P}_{\tau}\,=\,\vec{P}_{ 0} W^{\tau}.
\eeq{sinumintevolve}
\end{linenomath}
The abundance probability distribution conditioned against extinction and monodominance is 
\begin{linenomath}
\beq
P_{cn_1,\tau}\equiv\frac{P_{n_1,\tau}}{1-P_{0,\tau}-P_{J,\tau}},
\eeq{sinumintcond}
\end{linenomath}
for $n_1=1,\dots,J-1$.  

\subsection{Recovering non-zero-sum dynamics from zero-sum dynamics}
\label{zerosumtoordinary}

\noindent Our general approach is to treat empty space as the $(S+1)$th species in a community of $S$ species.  Let $w_{(S+1),0}$ be the rate at which death events generate empty space and set all the $a_{ij}$ to zero for $i,j=S+1$.  Given this setup, we consider the dynamics of a large--$J$ community as $n_{S+1}\rightarrow J$.  
\\
\\
\noindent Starting from the master equation for the Moran model in Eq.~\ref{nichezerosummaster}, we find
\begin{linenomath}
\beqa
T_{i,S+1,\vec{n}}&\sim& \frac{n_i}{J}, \nonumber \\
T_{S+1,i,\vec{n}}&\sim&  \frac{w_{i,\vec{n}}}{w_{(S+1),0}}\frac{n_i}{J},
\eeqa{sizstotijnexpand}
\end{linenomath}
and all other transition probabilities are higher-order in $n_i/J$ for $i \ne S+1$.  Now let $r$ be the overall transition rate.  Rescaling $w_{i,\vec{n}}\rightarrow w_{(S+1),0}w_{i,\vec{n}}/r$, setting $\tau=rJt$, and identifying $T_{S+1,i}$ and $T_{i,S+1}$ with $g_{i,\vec{n}}$ and $r_{i,\vec{n}}$, respectively, we find that Eq.~\ref{nichezerosummaster} reduces to Eq.~\ref{nichebdmaster} with $d_i=r$.  

Starting from the mean dynamics of the Moran model in Eq.~\ref{nichezerosumode}, and using $p_i=n_i/J$, we find
\begin{linenomath}
\beq
\frac{dn_i}{d\tau}\,\sim\, \frac{w_{i,\vec{n}}}{w_{(S+1),0}}\frac{n_i}{J}-\frac{n_i}{J},
\eeq{sizstomf}
\end{linenomath}
with all other terms being higher-order in $n_i/J$ for $i \ne S+1$.  The same rescalings as before yield Eq.~\ref{nichebdode} with $d_i=r$.

\subsection{Corrections to stability criteria of the simple Moran model for low levels of speciation and migration}
\label{correctionstostabilizationcriteria}

\noindent In a nearly neutral metacommunity where only the first species is distinct in ecological function, parameters for the symmetric species are identical:  $w_{i,0}=w_{2,0}$ and $a_{ij}=a_{2j}$ for all $i>1$ and all $j$.  If the number of symmetric species, $S-1$, is large such that terms of ${\cal O}(S)$ can be ignored in Eq.~\ref{metagenode}, then the mean dynamics for the asymmetric species can be written as
\begin{linenomath}
\beq
\frac{dp_1}{d\tau}\,=\,\frac{1}{J_M}\frac{(1-\nu_1)e^{-((B_1+B_2)p_1-B_2)}-(1-\nu_2)}{e^{-((B_1+B_2)p_1-B_2)}p_1+1-p_1}p_1(1-p_1)-\frac{\nu_2}{J_M}p_1,
\eeq{siscasymmode}
\end{linenomath}
where $B_1$ and $B_2$ are given by Eq.~\ref{metadeterbdefs} with the substitution $J\rightarrow J_M$.  The stable fixed point of the nearly neutral metacommunity can be calculated perturbatively in $\nu_1$ and $\nu_2$.  At leading order, we find
\begin{linenomath}
\beqa
p_1^*&=&\frac{B_2 -C_{\nu_1} \nu_1 - C_{\nu_2}\nu_2}{B_1+B_2}+{\cal O}(\nu_1^2,\nu_2^2,\nu_1 \nu_2), \nonumber \\
p_{i>1}^*&=&\frac{1-p_1^*}{S-1},
\eeqa{siscfp}
\end{linenomath}
where
\begin{linenomath}
\beqa
C_{\nu_1}&=&1, \nonumber \\
C_{\nu_2}&=&\frac{B_2}{B_1}.
\eeqa{siscfpcorrect}
\end{linenomath}
Stability requirements can be found from the linearization
\begin{linenomath}
\beq
\frac{dp_1}{d\tau}\,=\,-\frac{1}{J_M}\frac{(B_1+D_{\nu_1}\nu_1)(B_2+D_{\nu_2}\nu_2)}{B_1+B_2}(p_1-p_1^*)+{\cal O}(\nu_1^2,\nu_2^2,\nu_1 \nu_2),
\eeq{sisclin}
\end{linenomath}
where
\begin{linenomath}
\beqa
D_{\nu_1}&=& \frac{B_2^2-B_1^2-B_1B_2^2}{B_2(B_1+B_2)},  \nonumber \\
D_{\nu_2}&=&\frac{B_2^2}{B_1^2}\left(2-\frac{B_1B_2}{B_1+B_2}\right).
\eeqa{siscgrcorrect}
\end{linenomath}

\noindent For a nearly neutral local community, the dynamics of the asymmetric species, as prescribed by Eq.~\ref{localgenode}, can be written as
\begin{linenomath}
\beq
\frac{dp_1}{d\tau}\,=\,\frac{1}{J_M}\frac{(1-\nu_1)e^{-((B_1+B_2)p_1-B_2)}-(1-\nu_2)}{e^{-((B_1+B_2)p_1-B_2)}p_1+1-p_1}p_1(1-p_1)-\frac{\nu_2}{J_M}p_1,
\eeq{siscasymmodel}
\end{linenomath}
where $B_1$ and $B_2$ are given by Eq.~\ref{metadeterbdefs} with the substitution $J\rightarrow J_L$.  The stable fixed point of the nearly neutral local community can be calculated perturbatively in $m_1$ and $m_2$.  At leading order, we find
\begin{linenomath}
\beqa
p_1^*&=&\frac{B_2 -C_{m_1}m_1 - C_{m_2}m_2}{B_1+B_2}+{\cal O}(m_1^2,m_2^2,m_1 m_2), \nonumber \\
p_{i>1}^*&=& \frac{1-p_1^*}{S-1},
\eeqa{lsiscfplocal}
\end{linenomath}
where
\begin{linenomath}
\beqa
C_{m_1}&=&1-x_1\frac{B_1+B_2}{B_2}, \nonumber \\
C_{m_2}&=&\frac{B_2}{B_1}-x_1\frac{B_1+B_2}{B_1}.
\eeqa{siscfpcorrectlocal}
\end{linenomath}
Stability requirements can be found from the linearization
\begin{linenomath}
\beq
\frac{dp_1}{d\tau}\,=\,-\frac{1}{J_L}\frac{(B_1+D_{m_1}m_1)(B_2+D_{m_2}m_2)}{B_1+B_2}(p_1-p_1^*)+{\cal O}(m_1^2,m_2^2,m_1 m_2),
\eeq{sisclinl}
\end{linenomath}
where
\begin{linenomath}
\beqa
D_{m_1}&=&\frac{1}{B_2^2(B_1+B_2)}\left(B_2^3(1-x_1)+2B_1^2B_2^2x_1(1-x_1)\right. \nonumber \\
&&\left.-B_1B_2(B_1+B_2^2(1-x_1)^2-3B_1x_1)+B_1^3x_1(2-B_2x_1)\right), \nonumber \\
D_{m_2}&=&\frac{1}{B_1^2(B_1+B_2)}\left(2 B_2^3(1-x_1)+2 B_1^2 B_2^2 x_1(1-x_1)\right.  \nonumber \\
&&\left.+B_1 B_2^2(2-B_2(1-x_1)^2-3x_1)+B_1^3 x_1(1-B_2 x_1) \right).
\eeqa{sisclincorrect}
\end{linenomath}

\subsection{Calculation of extinction times in a nearly neutral metacommunity}
\label{extinctiontimes}

\noindent If  we assume a sufficiently large number of symmetric species such that ${\cal O}(1/S)$ terms in the master equation are negligible, marginal dynamics for the asymmetric species are governed by Eq.~\ref{asymmPnM} with
\begin{linenomath}
\beqa
g_{n_1}&=&\frac{J_M-n_1}{J_M}\left((1-\nu_1)\frac{e^{-((B_1+B_2)n_1/J_M-B_2-a_{12}/w_{1,0}+a_{22}/w_{2,0})} n_1}{e^{-((B_1+B_2)n_1/J_M-B_2-a_{12}/w_{1,0}+a_{22}/w_{2,0})} n_1+ J_M-n_1-1}\right), \nonumber \\
r_{n_1}&=&\frac{n_1}{J_M}\left((1-\nu_2)\frac{J_M-n_1}{e^{-((B_1+B_2)n_1/J_M-B_2+a_{21}/w_{2,0}-a_{11}/w_{1,0})} (n_1-1)+J_M-n_1}+\nu_2\right), \nonumber \\
\eeqa{metageasymmgrM}
\end{linenomath}
where $B_1$ and $B_2$ are given by Eq.~\ref{metadeterbdefs} with the substitution $J\rightarrow J_M$.  Hubbell's univariate metacommunity dynamics (Hubbell 2001) are included as the fully symmetric limit.  For an initial abundance of $n_{1,0}$, the mean times to extinction, $\tau_E$, are calculated using \citep[p.~260]{Gardiner:2004}
\begin{linenomath}
\beq
\tau_E\,=\,\sum_{p=0}^{n_{1,0}-1}\phi_p^M\sum_{q=p+1}^{J_M-1}\frac{1}{g_q^M\phi_q^M},
\eeq{metagente}
\end{linenomath}
where $\phi_0^M=1$ and for $p>0$
\begin{linenomath}
\beq
\phi_p^M\,=\,\prod_{m=1}^p\frac{r_m^M}{g_m^M}.
\eeq{metagenphidef}
\end{linenomath}

\subsection{Fluctuations in large local communities}
\label{localcv}

\noindent In Hubbell's theory of local communities, the expected abundance of each species at equilibrium is 
\begin{linenomath}
\beq
\lim_{\tau\rightarrow \infty}\langle N_i(\tau) \rangle\,=\,x_iJ_L.
\eeq{silcve}
\end{linenomath}
\citet{Vallade:2003p3851} first calculated the variance at equilibrium
\begin{linenomath}
\beq
\lim_{\tau\rightarrow \infty} \langle N_i(\tau) - \langle N_i(\tau) \rangle \rangle^2\,=\,x_i(1-x_i)J_L\frac{J_L+I}{1+I},
\eeq{silcvv}
\end{linenomath}
where $I=(J_L-1)m/(1-m)$ is called the ``fundamental dispersal number"~\citep{Etienne:2005p3829}.  Then, for large $J_L$, we obtain the approximation in Eq.~\ref{localdetercv}.

\subsection{A stationary distribution for the asymmetric species in a nearly neutral local community with weak competitive interactions}
\label{localnnstationary}

\noindent Marginal dynamics for the asymmetric species are governed by Eq.~\ref{asymmPnM} with
\begin{linenomath}
\beqa
g_{n_1}&=&\frac{J_L-n_1}{J_L}\left((1-m_1)\frac{\rho_g(n_1) n_1}{\rho_g(n_1) n_1+ J_L-n_1-1}+m_1\frac{\rho_g(n_1) x_1}{\rho_g(n_1) x_1+1-x_1}\right), \nonumber \\
r_{n_1}&=&\frac{n_1}{J_L}\left((1-m_2)\frac{J_L-n_1}{\rho_r(n_1) (n_1-1)+J_L-n_1}+m_2\frac{1-x_1}{\rho_r(n_1) x_1+1-x_1}\right), \nonumber \\
\eeqa{sisdasgr}
\end{linenomath}
where 
\begin{linenomath}
\beqa
\rho_g(n_1)&=&e^{-((B_1+B_2)n_1/J_L-B_2-a_{12}/w_{1,0}+a_{22}/w_{2,0})}, \nonumber \\
\rho_r(n_1)&=&e^{-((B_1+B_2)n_1/J_L-B_2+a_{21}/w_{2,0}-a_{11}/w_{1,0})}.
\eeqa{sisdasrhodef}
\end{linenomath}
and $B_1$ and $B_2$ are given by Eq.~\ref{metadeterbdefs} with the substitution $J\rightarrow J_L$.  We now assume weak competitive interactions such that 
\begin{linenomath}
\beqa
\rho_g(n_1)&\sim&1-((B_1+B_2)n_1/J_L-B_2-a_{12}/w_{1,0}+a_{22}/w_{2,0}) \nonumber \\
&\equiv&c_g+d n_1, \nonumber \\
\rho_r(n_1)&\sim&1-((B_1+B_2)n_1/J_L-B_2+a_{21}/w_{2,0}-a_{11}/w_{1,0})\nonumber \\
&\equiv&c_r+d n_1,
\eeqa{sisdasgrdef}
\end{linenomath}
where 
\begin{linenomath}
\beqa
c_g&=&1+\log\frac{w_{1,0}}{w_{2,0}}+(J_L-1)\frac{a_{22}w_{1,0}-a_{12}w_{2,0}}{w_{1,0}w_{2,0}}, \nonumber \\
c_r&=&1+\log\frac{w_{1,0}}{w_{2,0}}+J_L\frac{a_{22}w_{1,0}-a_{12}w_{2,0}}{w_{1,0}w_{2,0}}+\frac{a_{11}w_{2,0}-a_{21}w_{1,0}}{w_{1,0}w_{2,0}}, \nonumber \\
d&=&-\frac{w_{1,0}(a_{22}-a_{21})+w_{2,0}(a_{11}-a_{12})}{w_{1,0}w_{2,0}}. 
\eeqa{siscddef} 
\end{linenomath}
We specialize to the case where $a_{11}=a_{22}$, $a_{12}=a_{21}$, and $w_{1,0}=w_{2,0}$, so that $c_g=c_r\equiv c$.  The stationary distribution, $P_{n_1}^*\equiv\lim_{\tau\to\infty}P_{n_1}(\tau)$, is given by a well-known formula
\begin{linenomath}
\beq
P_{n_1}^* \,=\, P_0^* \prod_{i=0}^{n_1-1}\frac{g_{i}}{r_{i+1}}.
\eeq{sispstar}
\end{linenomath}
After some algebra, we obtain the closed form
\begin{linenomath}
\beqa
P^*_{n_1}&=&Z\binom{J_L}{n_1}\left(1+\frac{x(c+d n_1)}{1-x}\right)\eta^{n_1} (c/d)_{n_1}\nonumber \\
&&\times \frac{{\rm B}(\lambda_{a+}+n_1,\xi_{a+}-n_1){\rm B}(\lambda_{a-}+n_1,\xi_{a-}-n_1)}{{\rm B}(\lambda_{a+},\xi_{a+}){\rm B}(\lambda_{a-},\xi_{a-})} \nonumber \\
&&\times \frac{{\rm B}(\lambda_{b+}+n_1,\xi_{b+}-n_1){\rm B}(\lambda_{b-}+n_1,\xi_{b-}-n_1)}{{\rm B}(\lambda_{b+},\xi_{b+}){\rm B}(\lambda_{b-},\xi_{b-})} ,
\eeqa{sispstar1}
\end{linenomath}
where $(y)_z\equiv\Gamma(y+z)/\Gamma(y)$ is the Pochhammer symbol, $B(y,z)=\Gamma(y+z)/\Gamma(y)\Gamma(z)$ is the Beta function, and
\begin{linenomath}
\beqa
Z^{-1}&=&{}_6F_4(-J_L,c/d,\lambda_{a+},\lambda_{a-},\lambda_{b+},\lambda_{b-};1-\xi_{a+},1-\xi_{a-},1-\xi_{b+},1-\xi_{b-};-\eta) \nonumber \\
&&+ x c~{}_6F_4(-J_L,c/d+1,\lambda_{a+},\lambda_{a-},\lambda_{b+},\lambda_{b-};1-\xi_{a+},1-\xi_{a-},1-\xi_{b+},1-\xi_{b-};-\eta),\nonumber \\
\eeqa{sisz}
\end{linenomath}
and
\begin{linenomath}
\beqa
\lambda_{a\pm}&=& \frac{1}{2d}\left(c+d-1\pm\sqrt{(1-c+d)^2-4d(J_L-1)}\right), \nonumber \\
\lambda_{b\pm}&=&\frac{1}{2dx}\left(1-m-x+cx\pm\sqrt{(1-m-x+cx)^2-4mdx^2(J_L-1)}\right), \nonumber \\
\xi_{a\pm}&=&\frac{1}{2d}\left(1-c+2d\pm\sqrt{(1-c)^2-4d(J_L-1)}\right), \nonumber \\
\xi_{b\pm}&=&\frac{1}{2d(m_2-x)}\left.\bigg(1 + (d - c)m_2 - (d-c+1)x-(1-m_2) d x(J_L-1)\right. \nonumber \\
&&\left.\pm\sqrt{(1 - (d + c)m_2 + (d+c-1)x-(1-m_2) d x(J_L-1))^2} \cdots \right. \nonumber \\
&&\left.  \cdots\overline{-4(J_L-1) d (m_2 - x)(x+x(c+d)(m_2-1)-1)}\right.\bigg), \nonumber \\
\eta&=&\frac{dx}{m_2-x}.
\eeqa{sisotherdef}
\end{linenomath}
Eq.~\ref{sispstar1} is a generalized hypergeometric distribution~\citep{Kemp:1968tSUDDT,Johnson:1992tUDD}.

\clearpage 

\renewcommand{\thefigure}{\arabic{figure}}

\begin{figure*}
\vskip-2cm
\centerline{
\includegraphics[width=6cm]{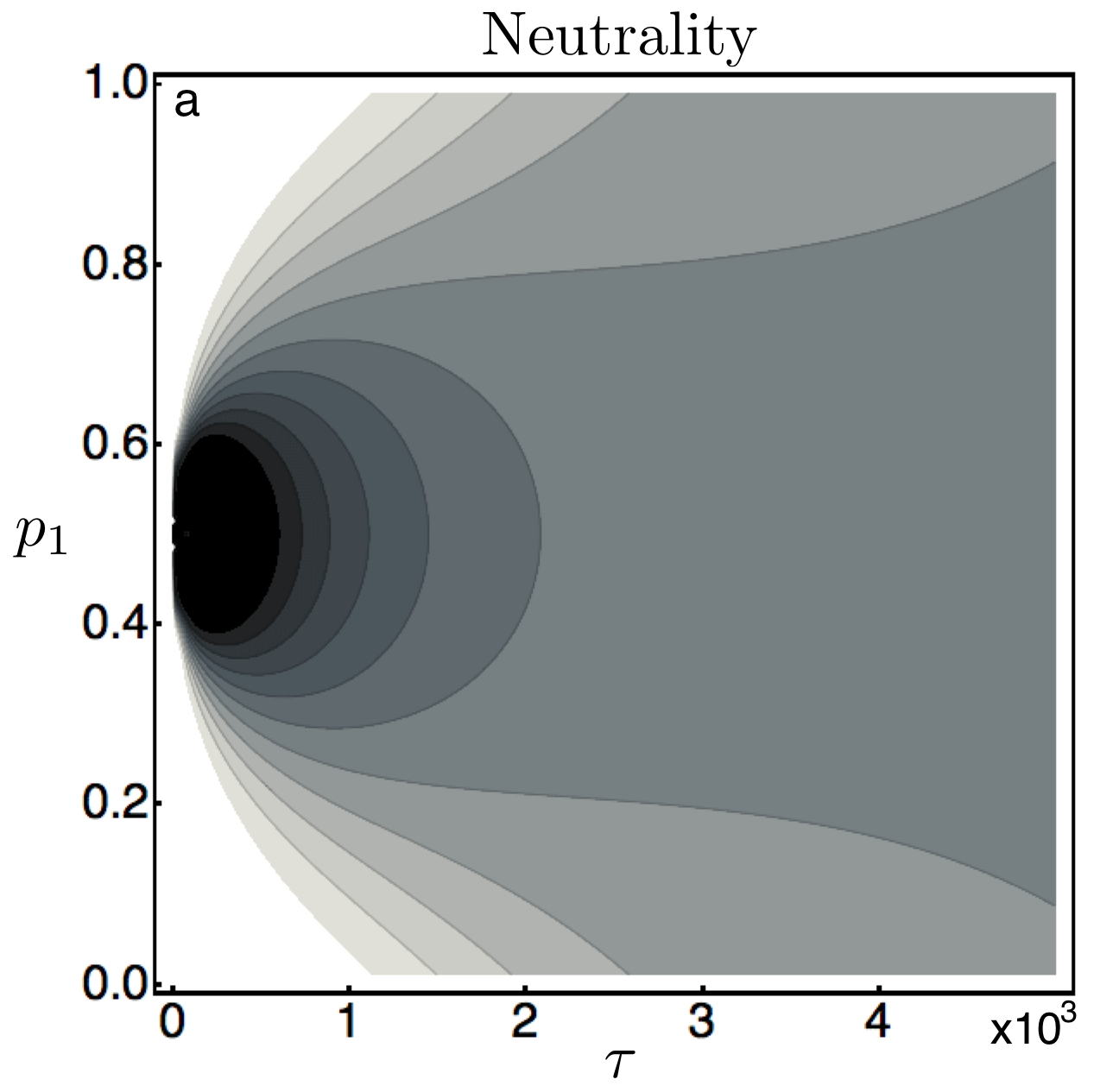}
\hskip0.00cm
\includegraphics[width=5.9cm]{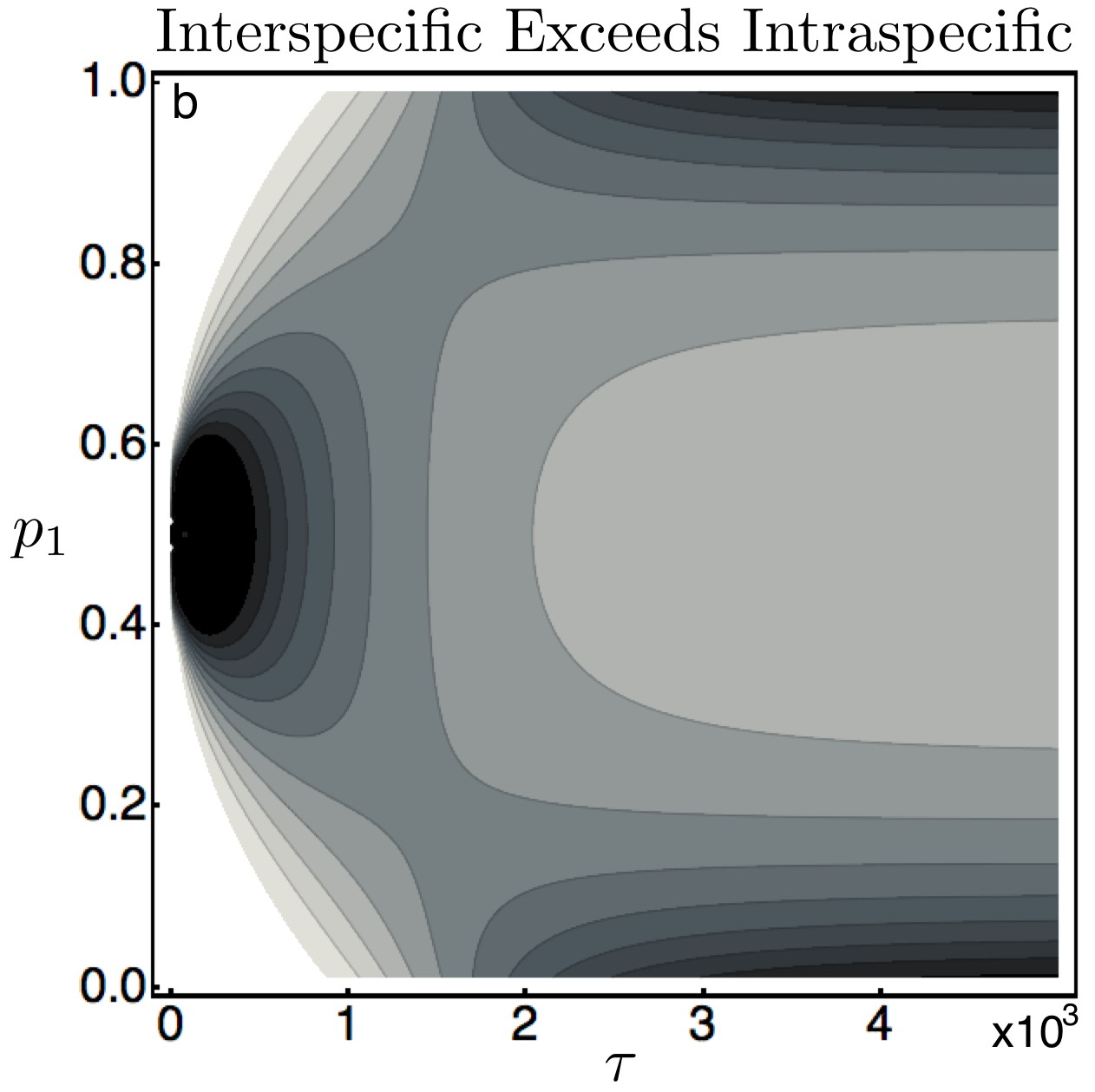}
}
\centerline{
\hskip0.00cm
\includegraphics[width=7.4cm]{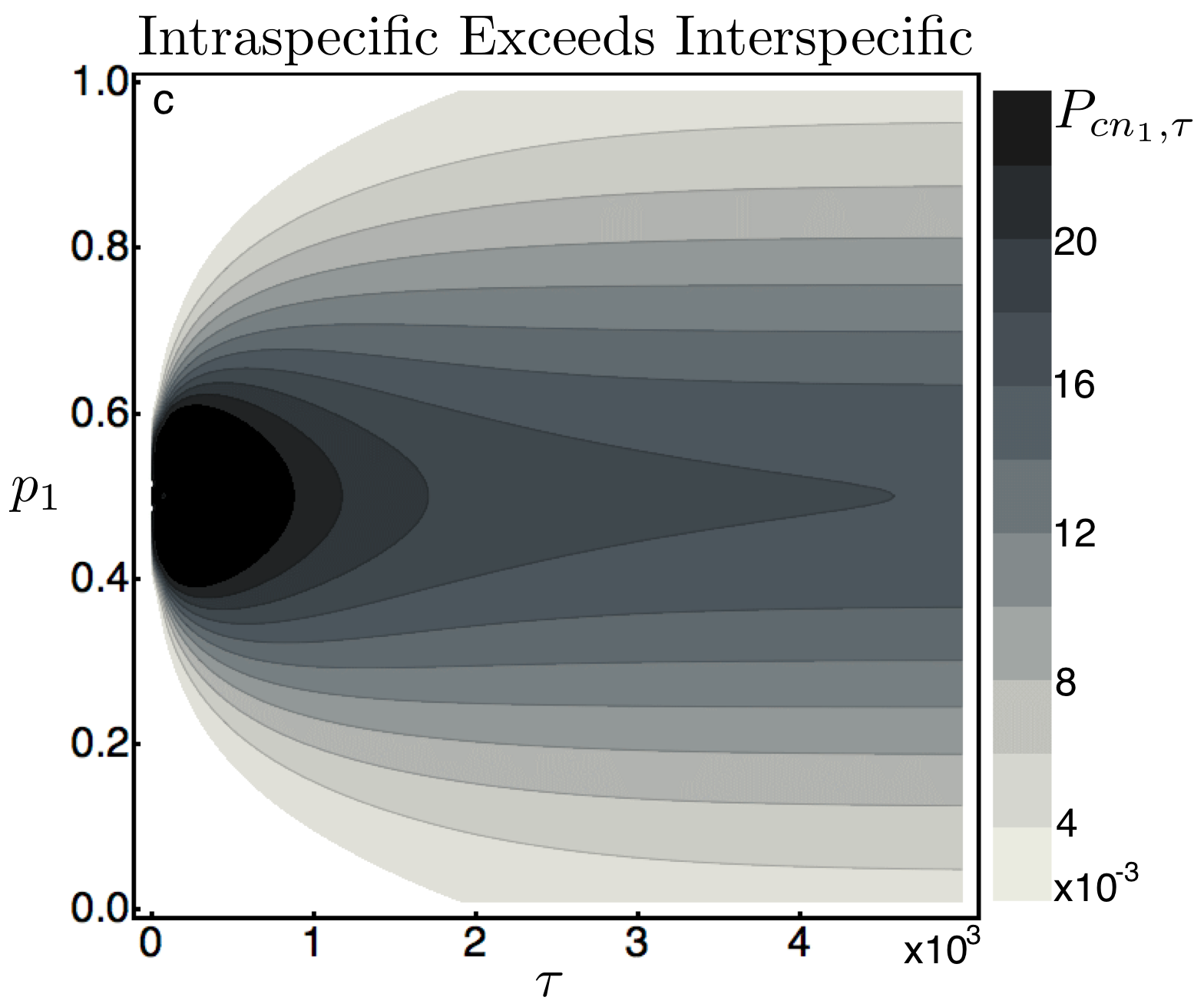}
}
\caption{Plots of the temporal evolution of conditional abundance probability distributions, $P_{cn_1,\tau}$ (see \ref{numericalintegration}).  For all plots, $J=100$ and $w_{1,0}=w_{2,0}$; the initial condition is $P_{cn_1,0}=\delta_{n_1,J/2}$; and the vertical axis is labelled by relative abundance, $p_1\equiv n_1/J$.  Panels a, b, and c correspond, respectively, to scenarios of neutrality ($a_{1j}=a_{2j}$ for $j=1,2$), interspecific exceeding intraspecific competition ($a_{11},a_{22}=0.1/J<a_{21},a_{12}=0.2/J$), and intraspecific exceeding interspecific competition ($a_{11},a_{22}=0.2/J>a_{21},a_{12}=0.1/J$).  In all cases, the conditional probabilities approach a quasi-stationary distribution at long times.  An excess of interspecific competition destabilizes the system, relative to the neutral scenario, by inducing a stochastic Allee effect and accelerating extinction for rare species.  By contrast, an excess of intraspecific competition generates a single peak in the quasi-stationary distribution and delays extinction.  Note that the peaks in $P_{cn_1,\tau}$ will approach delta functions in the $J\rightarrow\infty$ limit and the mean dynamics of the Levins niche model will be recovered in the absence of demographic stochasticity.  For finite $J$, absorbing states guarantee monodominance by a single species at long times, while niche dynamics accelerate or delay extinction on intermediate time scales.}
\label{moranstochintegration}
\end{figure*}

\begin{figure*}
\centerline{
\includegraphics[width=15cm]{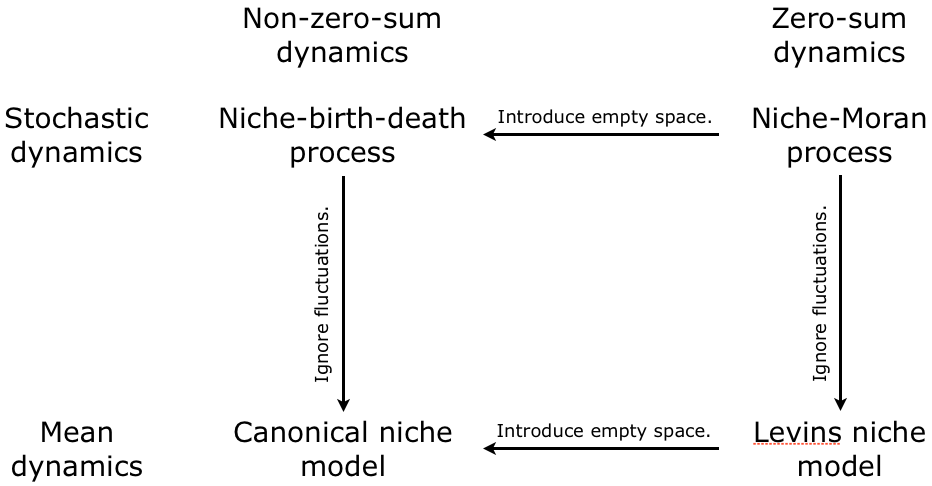}
}
\caption{A summary of connections among the blended stochastic processes and deterministic models, as discussed in the first part of this paper.  The second part develops non-neutral extensions of Hubbell's metacommunity and local community models by adding speciation and immigration, respectively, to the Niche-Moran process.  All models contain a neutral theory in the symmetric limit.}
\label{theorysumm}
\end{figure*}

\begin{figure*}
\vskip0.00cm
\centerline{
\includegraphics[width=7.5cm]{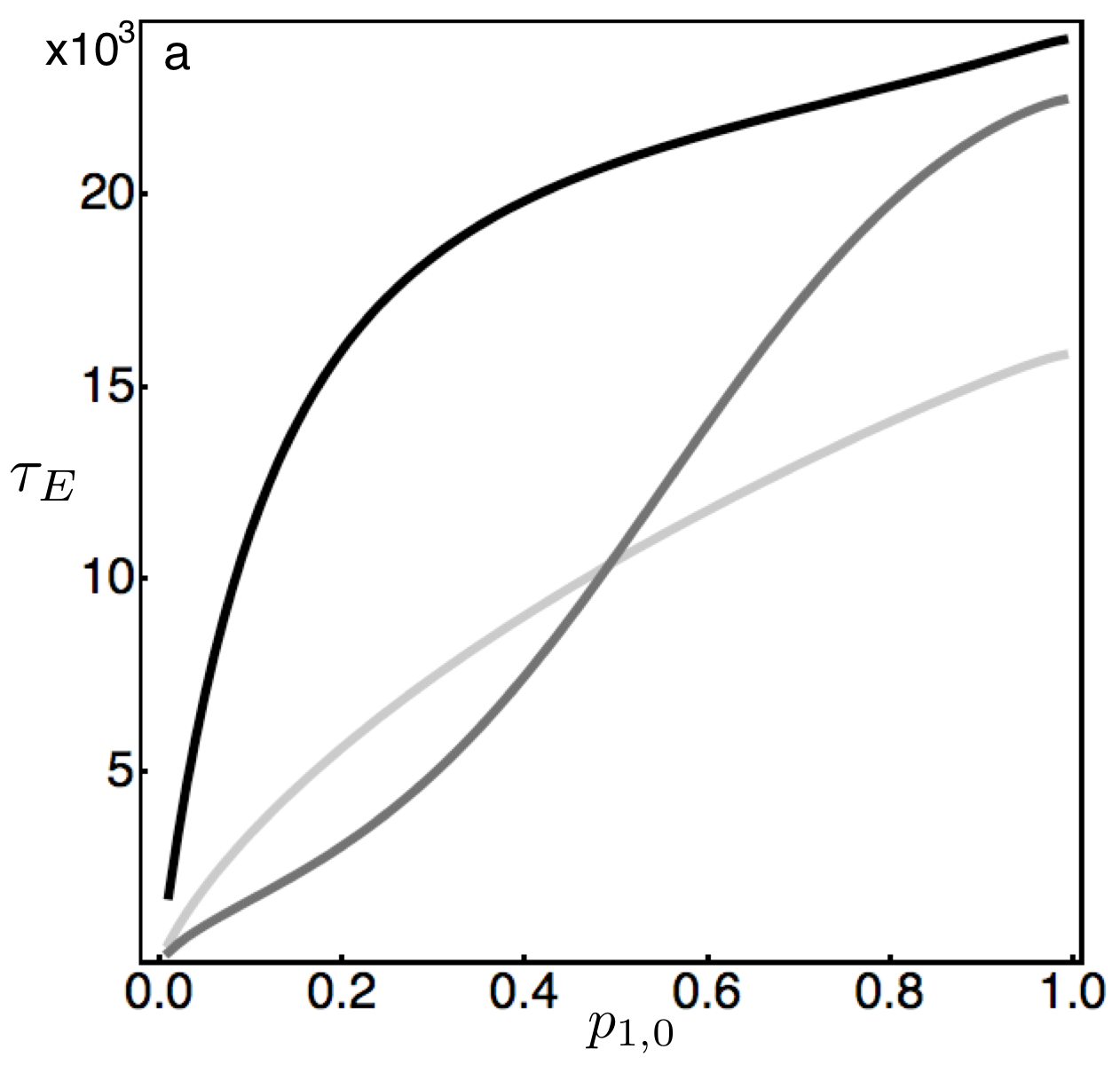}
\hskip0.00cm
\includegraphics[width=7.7cm]{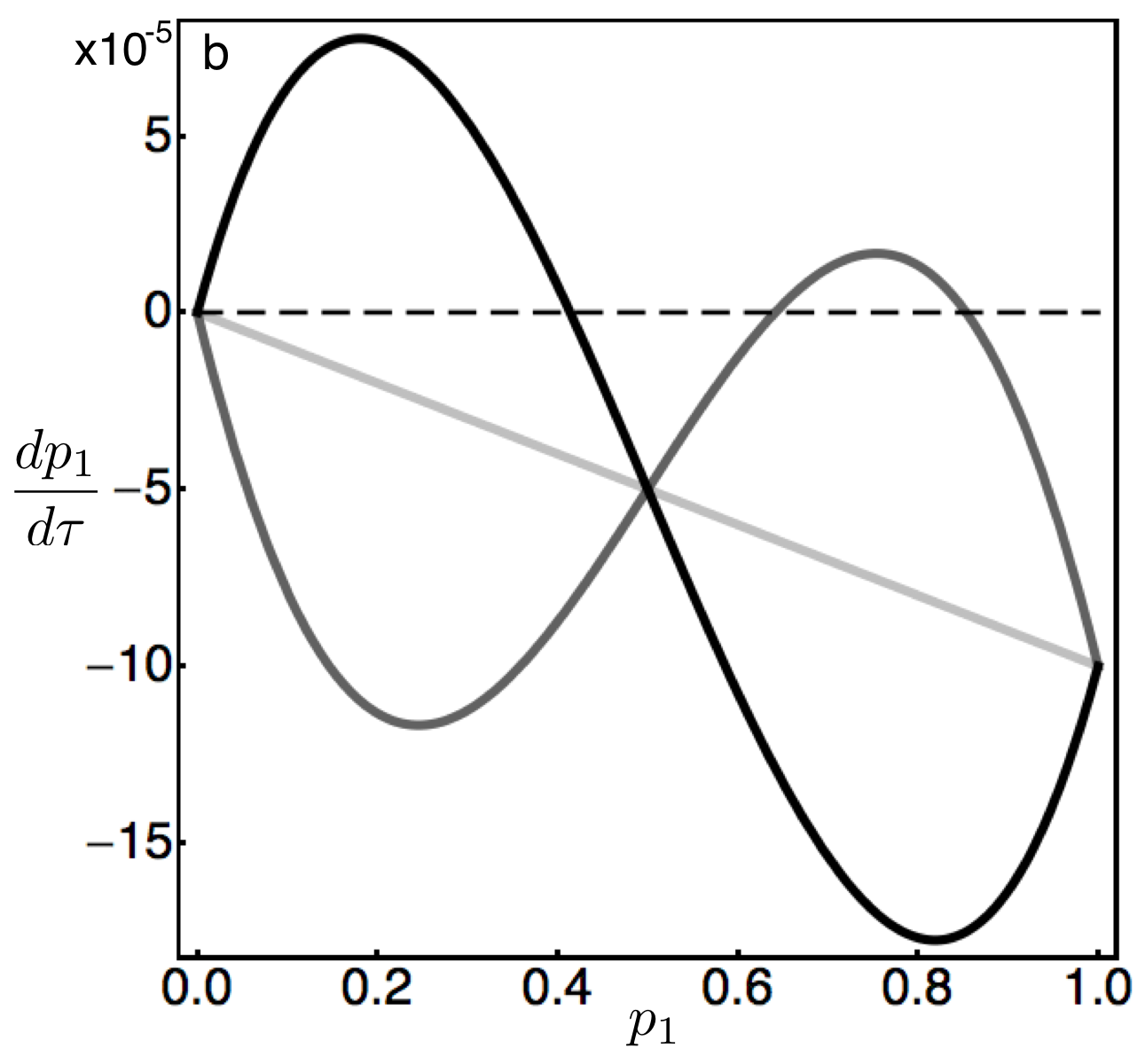}
}
\centerline{
\hskip0.80cm
\includegraphics[width=14cm]{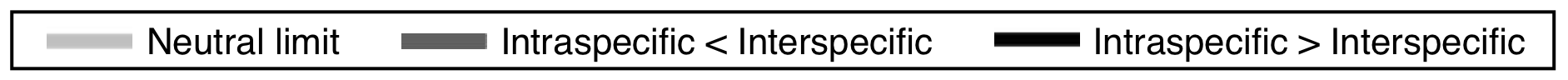}
}
\caption{Panel a plots mean times to extinction, $\tau_E$, against initial relative abundance, $p_{1,0} \equiv n_{1,0}/J_M$, for the asymmetric species in a nearly neutral metacommunity with $J_M=100$ individuals, while panel b plots the corresponding flows, $dp_1/d\tau$ versus $p_1$, of the mean dynamics.  We assume equivalence in speciation probability, with $\nu_1=\nu_2=0.01$, and intrinsic ecological fitness, such that $w_{1,0}=w_{2,0}$.  In the fully neutral limit, where $a_{1j}=a_{2j}$ for $j=1,2$, all species drift steadily toward extinction.  For the case of interspecific exceeding intraspecific competition, where $a_{11},a_{22}=0.1/J_M<a_{21},a_{12}=0.2/J_M$, mean time to extinction for the asymmetric species falls below (above) neutral expectations when relative abundance is low (high).  The existence of an inflection point in the $\tau_E$ versus $p_{1,0}$ curve signals a stochastic Allee effect, as confirmed by the mean dynamics where an unstable threshold separates an upper and a lower stable fixed point.  Only intraspecific exceeding interspecific competition, where $a_{11},a_{22}=0.2/J_M>a_{21},a_{12}=0.1/J_M$, raises $\tau_E$ above neutral expectations for all values of $p_{1,0}$.  The corresponding mean dynamics include a stable fixed point that promotes stable coexistence over intermediate time scales as well as an unstable fixed point that repels drift toward extinction.}
\label{metastochdeter}
\end{figure*}

\begin{figure*}
\vskip0.00cm
\centerline{
\includegraphics[width=8cm]{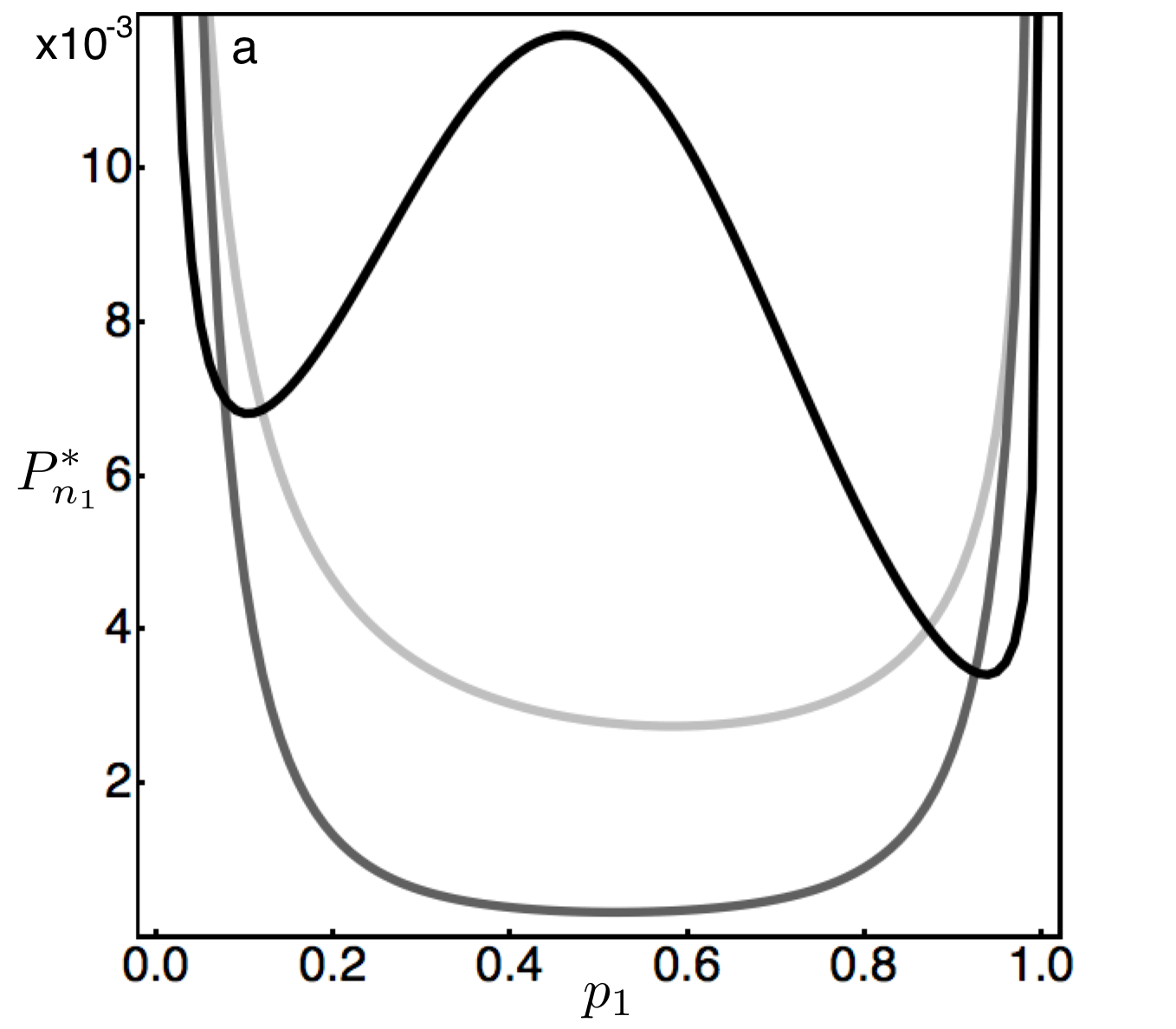}
\hskip0.00cm
\includegraphics[width=7.45cm]{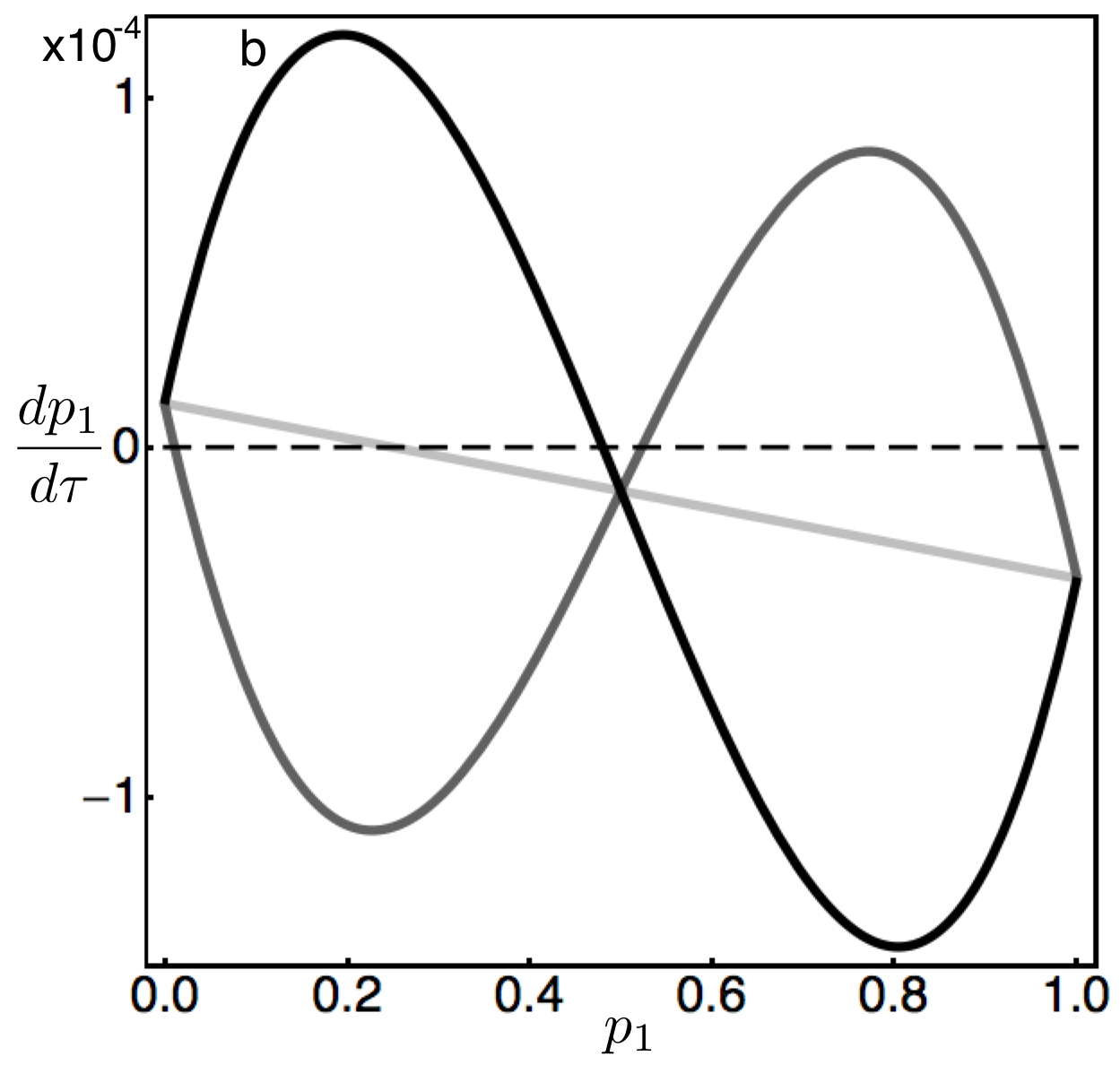}
}
\centerline{
\hskip0.80cm
\includegraphics[width=14cm]{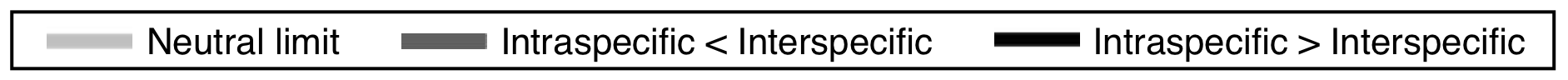}
}
\caption{Panel a plots the asymptotic (stationary) probabilities, the $P_{n_1}^*$, that the asymmetric species attains abundance $n_1$ in a nearly neutral local community of $J_L=100$ individuals.  Panel b plots the corresponding flows, curves of $dp_1/d\tau$, in the mean dynamics of the asymmetric species.  The relative metacommunity abundance is $x_1=0.25$; the migration probabilities are symmetric with $m_1=m_2=0.01$; and intrinsic ecological fitnesses are equivalent, such that $w_{1,0}=w_{2,0}$.  In the fully neutral limit, the stationary distribution is bimodal and large fluctuations between extirpation and monodominance destabilize the community.  Probabilities of coexistence decline further when interspecific exceeds intraspecific competition, such that $a_{11},a_{22}=0.1/J_M<a_{21},a_{12}=0.2/J_M$, and the mean dynamics exhibit an Allee effect.  By contrast, intraspecific exceeding interspecific competition such that $a_{11},a_{22}=0.2/J_M>a_{21},a_{12}=0.1/J_M$, generates a metastability at intermediate abundance that fosters coexistence over intermediate time scales and delays extirpation.  That metastability corresponds to a stable fixed point in the mean dynamics, where flows away from extirpation and monodominance further reduce the probability of large fluctuations.}
\label{localstochdeter}
\end{figure*}

\end{document}